\documentclass[reprint,amsmath,amssymb,aps,twocolumn,floatfix,showpacs,pre]{revtex4-1}

\usepackage{times}
\usepackage{amsmath}
\usepackage{amsfonts}
\usepackage{amssymb}
\usepackage{bm}
\usepackage{natbib}
\usepackage{graphicx}
\usepackage{latexsym}
\usepackage{enumerate}
\usepackage{hhline}

\numberwithin{equation}{section}
\usepackage[]{units}

\begin{document}

\title{Equilibrium self-assembly of small RNA viruses}

\author{R. F. Bruinsma$^{1,2}$, M. Comas-Garcia$^{3}$, R. F. Garmann$^{4}$, A. Y. Grosberg$^{5}$}
\affiliation{$^{1}$Department of Physics and Astronomy, University of California, Los Angeles, CA 90095,
USA}\affiliation{$^{2}$Department of Chemistry and Biochemistry, University of California, Los Angeles, CA 90095, USA}
\affiliation{$^{3}$HIV Dynamics and Replication Program, National Cancer Institute, Frederick National Laboratory for Cancer
Research,
Frederick, MD 21702}
\affiliation{$^{4}$School of Engineering and Applied Sciences, Harvard University, Cambridge, MA 02138, USA}
\affiliation{$^{5}$Department of Physics and Center for Soft Matter Research, New York University, 4 Washington Place, New
York, NY 10003}

\date{\today}

\begin{abstract}
We propose a description for the quasi-equilibrium self-assembly of small, single-stranded (ss) RNA viruses whose capsid
proteins (CPs) have flexible, positively charged, disordered tails that associate with the negatively charged RNA genome
molecules.  We describe the assembly of such viruses as the interplay between two coupled phase-transition like events:
the
formation of the protein shell (the capsid) by CPs and the condensation of a large ss viral RNA molecule. Electrostatic
repulsion between the CPs competes with attractive hydrophobic interactions and attractive interaction between neutralized
RNA segments mediated by the tail-groups. An assembly diagram is derived in terms of the strength of attractive
interactions between CPs and between CPs and the RNA molecules. It is compared with the results of recent studies of viral
assembly. We demonstrate that the conventional theory of self-assembly, which does describe the assembly of \textit{empty}
capsids, is in general not applicable to the self-assembly of RNA-encapsidating virions.
\end{abstract}

\pacs{87.15.bk, 87.15.kj}

\maketitle

\section{Introduction}

\subsection{Self-assembly of small RNA viruses}

\begin{figure}[htb]
\centering
\includegraphics[width=0.35\textwidth]{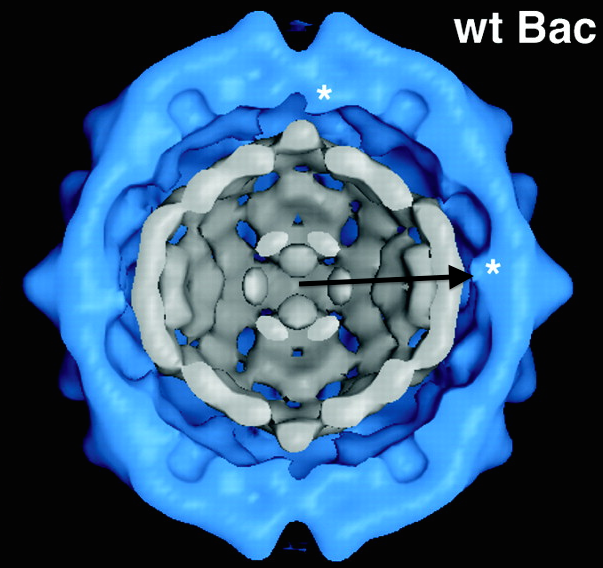}\\
\caption{(Color online) \label{fig:CCMV(1)} X-ray reconstruction of a cross-section of a $T=3$ virus-like particle (from ref.
\cite{Tihova}). The capsid is composed of 180 identical copies of Flock House Virus capsid proteins (``wild-type'' or wt)
arranged in an icosahedral shell (outer layer). The encapsidated RNA material is non-viral, so there are no specific
protein-RNA interactions. Only the part of the RNA material that has icosahedral symmetry is shown. The radius of
the condensed RNA globule, indicated by an arrow, is about \unit[10]{nm}. The stars indicate two-fold close contacts between
the enclosed RNA globule and the capsid.  The image is reproduced, with permission, from ref. \cite{Tihova} [copyright
(2004)
American Society of Microbiology.].}
\end{figure}

Assembly is a key part of the life cycle of a virus. During assembly, structural proteins and genome molecules produced
inside an infected cell combine to form virus particles (``virions''). Remarkably, many small viruses with a single-stranded
(ss) RNA genome (``vRNA'') will assemble under laboratory conditions in solutions that contain the protein and genome
molecular components of the virus \cite{Bancroft_1968, Adolph_1976}. Figure~\ref{fig:CCMV(1)} shows a reconstruction
\cite{Tihova} of the Flock House Virus (FHV), an example of a small ssRNA virus \footnote{The virus-like particle was
produced in a cell expression system (``wt-Bac'') that produces viral proteins, which encapsidate cellular RNA molecules}.
The icosahedral shell, or
``capsid'', has an inner radius $R_c$ of about \unit[10]{nm} and a thickness of about \unit[3]{nm}. It is composed of $180$
identical proteins (CPs). In the Caspar-Klug classification of viral capsids, icosahedral shells composed of 180 subunits
are
known as ``$T=3$'' shells.

The genome of a $T=3$ virus encodes minimally two proteins: the capsid protein and an RNA-dependent RNA polymerase, together
about 4,000 bases. A \unit[10]{nm} radius spherical volume can accommodate about 5,000 RNA bases in the form of a (hydrated)
crystal of duplex RNA. The density of the minimal RNA genome is thus not far below that of the hydrated RNA crystal form (in
some cases the density of the packaged RNA material even \textit{exceeds} that of the crystal form \cite{Rueckert,
Schneemann}). The dimensions of ss RNA genome molecules in solution are hard to measure, but the combined evidence from
small-angle x-ray scattering, cryo-EM, and fluorescence measurements indicates that vRNA molecules are swollen in
physiological solutions. For example, the hydrodynamic radius of the vRNA molecules of the MS2 virus has been estimated to
be
about \unit[14]{nm} \cite{Borodavka} whereas $R_c$ is about \unit[11]{nm} for MS2. Genome encapsidation thus requires a
significant level of compression of the vRNA molecules \footnote{The compression of the double-stranded DNA genome molecules
of phage viruses is carried out by a
molecular motor embedded in the viral capsid, but that is not the case for viruses with single-stranded RNA or DNA genomes}.

Viral self-assembly is driven by the competition between repulsive and attractive macromolecular interactions. It is well
known that specific affinities -- which involve stem-loop and tRNA-like motifs of the native vRNA molecules that bind
preferentially to the viral CPs in question -- significantly speed up assembly kinetics and allow for assembly at lower
concentrations of the components. Nevertheless, self-assembly studies of CPs with non-native RNA molecules (e.g.,
\cite{Cadena-Nava_2012}) indicate that generic interactions are in general capable of packaging ssRNA molecules in the
absence of specific CP-RNA affinities, albeit with reduced yield. For example, the virus-like particle shown in Fig.
\ref{fig:CCMV(1)} packages non-native ssRNA material. This article will focus exclusively on viral self-assembly driven by
generic interactions.

\begin{figure}[htb]
\centering
\includegraphics[width=0.30\textwidth]{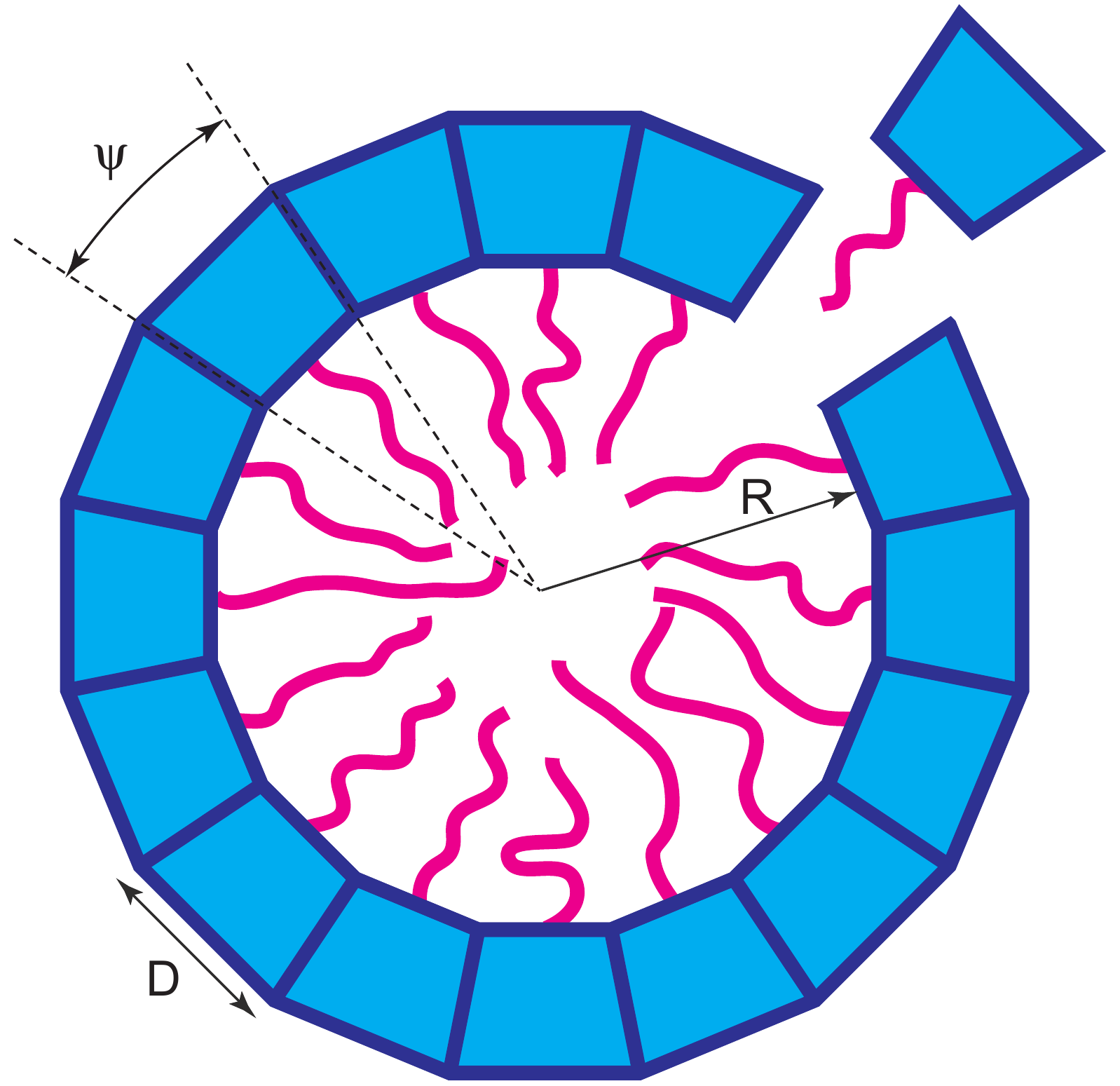}\\
\caption{(Color online) \label{fig:CCMV} Schematic cross-section of a small ssRNA viral capsid. The positively charged tail
groups of the capsid proteins extend inward where they can associate with sections of the negatively charged branched RNA
molecules (not shown). The angle $\psi$ is the relative angle between the normals of adjacent capsid proteins. For an inner
radius $R$ of about \unit[10]{nm} and a characteristic capsid protein (CP) dimension $D$ of about \unit[3]{nm} $\psi\simeq
D/R\approx 0.3$ radians. The figure is roughly to scale for a $T=3$ virus.}
\end{figure}

Among the generic interactions, \textit{electrostatics} plays a central role. The CPs of $T=3$ ssRNA viruses typically --  but
not always -- have a negatively charged ``head group'' and a positively charged ``tail group'' (see Fig. \ref{fig:CCMV}). The
pH-dependent negative charge $-eZ_h$ of the head group is located mostly on the part of the CP that faces the capsid
exterior, while the pH-independent positive charge $+eZ_t$ of the tail group faces the capsid interior. Typically, $Z_h\sim
Z_t\sim10$. The net charge of the CPs of the cowpea chlorotic mottle virus (CCMV)  --  a self-assembling $T=3$ ssRNA virus
whose assembly process
has been particularly well-studied -- is negative under physiological conditions but if the pH is reduced then the sign
changes around pH $\simeq 3.6$ \cite{Vega}, the isoelectric point \footnote{The positive tail group charge derives mainly
from arginine and lysine residues. It is not pH dependent because arginine and lysine do not titrate over relevant pH
values.}. The charge distribution of CPs that are part of a capsid has a dipolar component that remains very large -- of the
order of $\unit[10^3]{Debye}$ -- even at the isoelectric point. If the characteristic energy scale of the electrostatic
repulsion between CPs in a \unit[0.1]{M} salt environment is estimated by Debye-H\"{u}ckel (DH) theory, then values in the
range of $10 k_B T$ or more are found.

The positively charged CP tail groups have an electrostatic affinity for the negatively charged RNA nucleotides. Evidence is
provided by the fact that the strength of the affinity varies inversely with the ionic strength of the solution
\cite{Garmann}. Measured dissociation constants \cite{Bayer} for CP/RNA association binding give binding energies in the
range of $15 k_BT$. It should be noted that the CP/vRNA binding affinity can have important contributions coming from
correlation effects \cite{Boris(2)}. Numerical simulations \cite{Garcia, Devkota} report that the electrostatic affinity
involves \textit{counterion release}. The importance of CP-RNA electrostatic interactions is manifested by the fact that the
amount of vRNA that is packaged by a small ssRNA virus is a linear function of the net positive tail charge \cite{Belyi}.

Importantly, neutralization of the ssRNA material by the positive CP tail charges is \textit{incomplete}: the interior of a
CCMV virion has a large residual macroion charge in the range of $-10^3e$. This disparity between the CP
tail charge and the vRNA charge is a form of ``overcharging'', a fundamental issue in the theory of the electrostatics of of macroions \cite{PhysToday, Yan, Shura, Andrey, ZhangShklovskii_disproportionation}. Deviations from macroion charge neutrality in aqueous solutions are attributed to constraints and/or correlations that prevent matching of the positive and negative charges. In the context of viral assembly, overcharging was attributed by Hu \textit{et al.} \cite{Boris(2)} to the structure of the CP tail group/RNA association and to Manning condensation by Belyi \textit{et al.} \cite{Belyi}.

The repulsive electrostatic interactions between CPs, which inhibit capsid assembly, compete with highly directional CP-CP
``pairing'' attraction \cite{Kegel}. This attraction is provided by a combination of attraction between complementary
hydrophobic patches across CP-CP interfaces and pH-dependent proton-mediated pairing interactions between carboxylate groups
on residues of adjacent CPs facing each other (``Caspar pairing'' \cite{Caspar}). For capsid assembly to take place, the
strength of the attractive interactions between CPs must exceed that of repulsive electrostatic interactions, so it should
also be in the range of $10 k_BT$. The competition between the attractive pairing interactions with the salt and
pH-dependent
electrostatic repulsion is illustrated by  assembly diagrams of aqueous solutions of CCMV CPs (but no RNA) with the pH and
ionic strength levels as the thermodynamic variables \cite{Adolph-Butler, Lavelle}. Under conditions of high ionic strength
and reduced pH -- which means increased attraction -- empty capsids assemble spontaneously. This does not happen under
conditions of neutral acidity when the negative charge of the CP head groups apparently is just large enough to overcome the
attractive interactions.

The assembly of RNA-containing virions is normally described in terms of \textit{kinetic pathways} (e.g. \cite{Endres}).
Numerical studies of simple models for virion assembly \cite{Hagan3} show two distinct pathways. If the net CP-CP attraction
is large compared to the CP-RNA binding energy then assembly  proceeds via a classical nucleation-and-growth pathway. If the
net CP-CP attraction is small compared to the CP-RNA binding energy then assembly proceeds via a form of collective
condensation (``en masse'').

\subsection{Equilibrium self-assembly of virions}

This article was motivated by a recent series of self-assembly experiments of CCMV virions carried out under a protocol that
maintained, as closely as possible, thermodynamic equilibrium during assembly (e.g. \cite{Cadena-Nava_2012}). Equilibrium
thermodynamics has already been extensively applied to the self-assembly of \textit{empty} capsids \cite{Johnson} (see also
Supplemental Material, Sec. IV \footnote{For mathematical details and notation used in this work, see Supplemental Material at http://link.aps.org/supplemental/DOI.}). According
to equilibrium thermodynamics, the onset of capsid assembly as a function of the solution concentrations of the molecular
components has the character of a \textit{phase transition} in the limit that the number of molecular components per
aggregate is large compared to one. The critical CP concentration for this transition, sometimes called the ``critical
micelle concentration (CMC)'' \cite{McPherson2} by analogy with the self-assembly of micelles, is determined by the
condition
that the chemical potential of a CP in solution is the same as that of a CP that is part of a capsid. The predictions of
equilibrium thermodynamics agree well with chromatography studies of the self-assembly of empty capsids of CCMV
\cite{Johnson} and other other viruses. The characteristic energy scale for the CP-CP interactions in a CCMV capsid at
neutral pH was in the range of just a few $k_BT$, indicating that the repulsive electrostatic interactions between CPs
indeed
are closely balanced by the attractive interactions.

Under conditions of neutral pH and physiological salt concentrations, empty CCMV capsids do not form in solutions of CCMV
CPs
but addition of vRNA molecules leads to virion assembly when the pH is reduced \cite{Cadena-Nava_2012}. This stabilization
of
assembly by the vRNA molecules would seem to be obvious on the basis of straightforward electrostatic considerations: if the
positively charged tail groups of the CPs are neutralized by the negatively charged vRNA molecules then this should reduce
the electrostatic repulsion with respect to the hydrophobic attraction, and hence trigger assembly.
For CCMV at least, this argument is invalid. CCMV CPs whose tails have been removed do \textit{not} assemble under
conditions
of neutral pH \cite{Bancroft}. If neutralization of the tail groups was a sufficient condition for assembly then this should
have happened. Next, we already saw that CCMV CPs have a dipolar charge distribution and association of the tail group of a
CCMV CP with an ssRNA molecule actually \textit{increases} the total negative charge of the assembly (since $Z_h$ exceeds
$Z_t$ at neutral pH) and this strengthens electrostatics repulsion. At the isoelectric point, where $Z_h=Z_t$,  two CPs can
crudely be treated as oriented electrostatic dipoles. In that case, neutralization of the positive charges of the dipoles
\textit{still} increases the net repulsion between two CPs for larger separations. An additional source of attractive
interactions clearly is required for vRNA-triggered self-assembly of CCMV virions. This additional source of attraction will
be assumed to be the \textit{condensation} of vRNA molecules induced by the CP tail groups, as we will now discuss.

\subsection{Condensation of single-stranded nucleotide chains}

It is well known that double-stranded (ds) $\lambda$ phage B-DNA molecules in aqueous saline solutions condense into
rodlike
and toroidal aggregates when low concentrations of condensing agents are added to the solution \cite{Bloomfield, PhysToday}.
The condensing agents -- which can be neutral or positively charged polyvalent ions -- generate an effective short-range
attraction between ds DNA molecules \cite{Niels}.  When low concentrations of poly-L-Lysine are added to solutions
containing
plasmid length \textit{single-stranded} DNA molecules then condensation is observed as well \cite{Molas, Santai}. These
condensates are in fact, under the same conditions, significantly smaller and more stable then their dsDNA counterparts.
They
have a disordered, spherical appearance under TEM \cite{Molas, Santai} and they tend to aggregate together. Which of these
two different modes of condensation prevails is believed to be determined by the \textit{persistence length}. The
persistence
length of ssDNA chains is roughly a factor $50$ smaller than that of ds chains. Numerical simulations of linear
homopolymers with self-attraction report that, with increasing persistence length,  a structural transformation takes place
in the morphology of the condensates from a disordered, spherical globule to an ordered toroidal condensate  \cite{Noguchi,
Ou}.

Disordered spherical condensates appear in solutions of flexible, charged polymers (``polyelectrolytes'') to which
polyvalent
ions have been added as condensing agents. In polymer physics the appearance of such condensates are viewed as an example of
the ``coil-to-globule'' transition \cite{RedBook}. In this article, we will assume that swollen ss vRNA molecules in
solution
condense via a coil-to-globule transition when condensing agents are added and that CPs in general, and the tail-groups in
particular, act as the vRNA condensing agents.

\subsection{Equilibrium assembly diagram}

\begin{figure}[htb]
\centering
\includegraphics[width=0.40\textwidth]{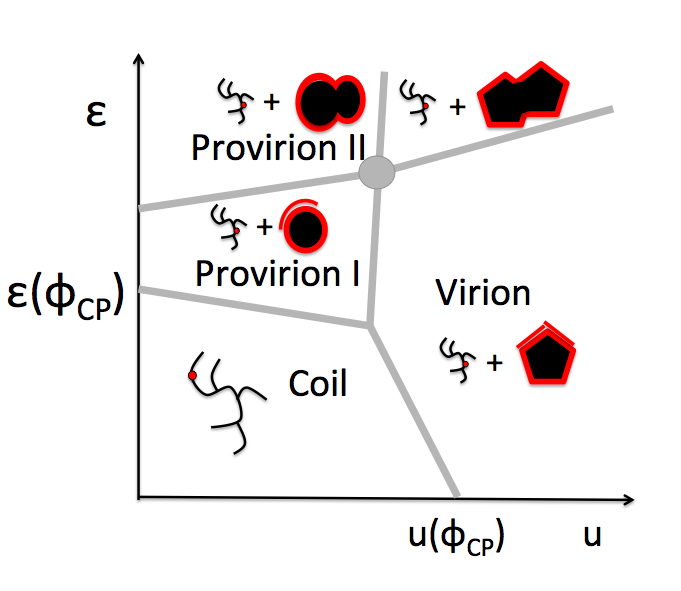}\\
\caption{(Color online) \label{fig:phase} Equilibrium assembly diagram of a small RNA virus. The vertical axis $\varepsilon$
is the binding energy of a capsid protein (CP) to the viral vRNA molecule. The horizontal axis $u$ is the strength of
attractive CP/CP pairing interactions. Progressive pH reduction at fixed salinity roughly corresponds to a horizontal path in
this diagram.
Schematic boundaries between the different regimes, indicated in light gray, are a guide to the eye only.}
\end{figure}

\begin{figure}[htb1]
\centering
\includegraphics[width=0.7\columnwidth]{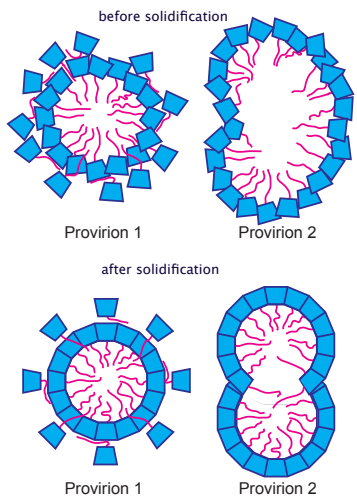}\\
\caption{(Color online) \label{Provirions} Provirion 1 and 2 states before (top) and after (bottom) solidification. The
number of capsid proteins (CPs) in the first layer of the provirion 1 state whose tails are strongly associated with the RNA
molecule(s) is comparable to that of the virion. The interior is negatively charged.  The tails of the excess CPs in the
second layer have a much weaker association with RNA, either due to the interaction of their tails with the outer side of
the
first CP layer, or because their tails are forced to squeeze in between the CPs of the first layer to access the RNA. In the
$provirion 2$ state, the number of CPs that are strongly associated with RNA is larger than that of a virion. The tail groups
fully neutralize the RNA molecule(s). }\end{figure}

Based on the model discussed in the next sections, a schematic equilibrium assembly diagram is obtained shown in
Fig.\ref{fig:phase}.
The vertical axis $\varepsilon$ is the binding affinity between a CP and a vRNA molecule -- including correlation effects
\cite{Boris(2)} -- while the horizontal axis $u$ is the strength of attractive CP/CP pairing interactions. If both of these
parameters are small compared to $k_BT$, then the vRNA molecules are swollen and most CPs are free in solution (not shown).
Increasing $\varepsilon$ increases the number of CPs associated with a vRNA molecule, which effectively reduces the solvent
quality. At a threshold $\varepsilon(\phi_{\mathrm{CP}})$ -- which depends on the total CP concentration $\phi_{\mathrm{CP}}$
-- the solution \textit{disproportionates} into condensed CP-rich ``saturated aggregates'' and swollen CP-poor vRNA
molecules.
Disproportionation is similar to phase separation but without the appearance of phase boundaries
\cite{Kabanov_disproportionation, ZhangShklovskii_disproportionation}. Instead, the solution is a uniform mixture of two
different populations of aggregate species in thermal equilibrium with each other, as further discussed in Sect. II D. In
the present case, one species is composed of swollen vRNA molecules with a small number of associated CPs while the other
species is composed of aggregates of condensed vRNA molecules surrounded by a layer of headgroups in a liquidlike state
(the $provirion 1$ state in Fig.\ref{Provirions}). The CPs are forced out of the interior of a condensed vRNA globule by a
combination of \textit{surface tension} of the globule and angle-dependent pairing attraction between the CPs.

 In the provirion 1 state, the condensed, spherical vRNA molecule is surrounded by (roughly) 180 CPs whose tailgroups are
 associated with the vRNA molecule. The tail groups do not neutralize the vRNA molecule so the interior has a net negative
 macroion charge. Excess CPs either are free in solution or physisorbed on the surface of the provirion. The two populations
 of bound and free CPs are in thermal equilibrium with each other.  If the strength $u$ of the pairing attraction between
 the
 CPs is increased then the CPs of the inner layer of the provirion 1 state crystallize out into a $T=3$ shell of exactly 180
 CPs (see Fig.4). If $\varepsilon$ is increased for fixed (small) $u$, then a second transition is encountered beyond which
 the CP tail groups
neutralize the vRNA molecule. This is the provirion 2 state, which has a significantly larger surface area than
the provirion 1 state but a similar volume. It has zero surface tension and is subject to shape fluctuations. Solidification
starting from the provirion 2 state is expected to lead to ``malformed shells'' composed of more than 180 CPs (see Fig.4).
The
existence of a provirion 2 state is one of the central predictions of the theory. Provirion 2 aggregates would be a novel
application area for the physics of strongly fluctuating interfaces \cite{Safran}, developed originally for surfaces and
interfaces composed of amphiphilic molecules. However, the action of the CPs is not due to the competition between the
hydrophobic and hydrophilic parts of an amphiphilic molecule but due to the affinity of the positively charged tail-groups
for the interior of the condensed vRNA molecules and of the negatively charged head-groups for the exterior. If provirion
2 particles are indeed found then one might say that CPs act as ``amphielectrics''.

An essential claim of the proposed model is that virion assembly for larger values of $\varepsilon$ does {\bf{not}} follow
the conventional theory of
self-assembly \cite{Safran} and the Law of Mass Action (see also Supplemental Material, Sect. IV [30]) of equilibrium chemical
thermodynamics. The key point of the model is that virion assembly will take place from a pre-condensed CP/vRNA aggregate for
sufficiently large $\varepsilon$. The CP concentration inside such an aggregate can be very high, even when the CP solution
concentration is very low, provided that
$\varepsilon$ is large enough to off-set the low CP solution chemical potential and allow aggregate formation. The final
assembly of the virion from this pre-condensed state for increasing $u$ is then independent of the solution CP concentration.
In essence, the aggregate acts as a chemical reactor that concentrates the components.

If the strength $u$ of the attractive interactions is increased for $\varepsilon<\varepsilon(\phi_{\mathrm{CP}})$, then
there
is no disproportionation. Instead, virion assembly takes place directly from the swollen coil phase and follows the
conventional Law of Mass Action scenario of empty capsid assembly . We speculate -- but have not shown -- that in terms of
assembly kinetics, a direct transition from the coil phase to the virion phase for lower $\varepsilon$ obeys the nucleation
and growth scenario. In contrast, virion assembly starting from the provirion 1 and 2 states -- so for larger $\varepsilon$ --
is expected to proceed via some form of the ``en masse'' kinetic scenario. An interesting aspect of the phase diagram of
Fig.\ref{fig:phase} is the fact that the equilibrium phase diagram includes the provirion I structure as a stable structure.
If an assembly experiment would produce a structure like the provorion I, then this would normally be
interpreted as a ``kinetic trap''. That does not mean that there are no kinetically trapped states in the proposed model. If
an equilibrium phase diagram includes multiple competing structures separated by first-order transition lines -- as is the
case for the proposed model -- then this only enhances kinetic trapping. The equilibrium assembly of virions requires in
general a very fine balance between competing nonspecific interactions.

In Sect. II, we present the simplest version of the model that describes the condensation of vRNA molecules as a
coil-to-globule transition induced by CPs. Section III extends the model to include capsid formation deeper in the condensed
phase. In the concluding Sect. IV we compare the model with the recent experiments on the equilibrium assembly of CCMV,
and discuss experiments that would help to verify (or disprove) the model. We conclude with a discussion of the limitations of
the model and how it could be extended further. For the convenience of the reader, a table of symbols used in the paper is
provided in the Supplemental Material [30].

\section{Coil-to-globule transition}
In its simplest form, the model is a variational free energy for a homogeneous CP/RNA aggregate in terms of the radius of
gyration $R$ of the aggregate, the maximum ladder distance $S$ (or \textit{MLD}), defined as the maximum number of
complementary paired nucleotides separating two points of the RNA molecules \cite{Fang}, and the segment occupation
probability $x$. The latter is defined as the probability that a segment of the vRNA molecule is associated with the tail
group of a CP. The variational free energy $F(R,S,x)$ is defined as

\begin{equation}\label{fmix}
\begin{split}
&{\beta F(R,S,x)} =\frac{R^2}{l^2S}+\frac{S^2}{N}+ V(x)\frac{N^2}{R^3}+W\frac{N^3}{R^6}-\\ &-Nx\beta\varepsilon+N \left[
x\ln
x+(1-x)\ln(1-x)\right]+\beta F_{PB}(R) \quad\quad  .
\end{split}
\end{equation}
with $\beta=1/k_BT$. The different terms will be explained in sequence.

\subsection{Coil-to-globule transition of annealed branched polymers}

\begin{figure}[htb]
\centering
\includegraphics[width=0.2\textwidth]{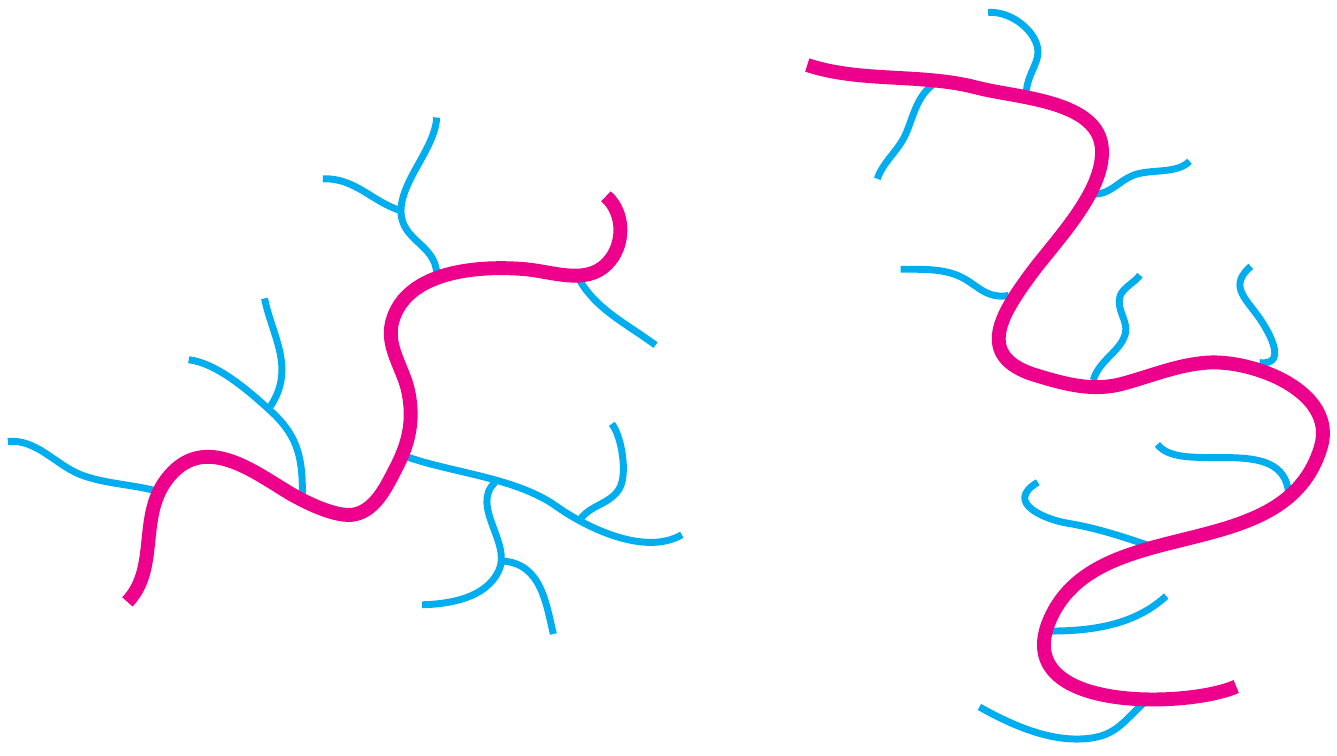}\\
\caption{(Color online) \label{Toy2} Different realizations of branched tree structures composed of $N$=21 segments. The first
structure has a maximum ladder distance $S$=6 while $S$=11 for the second structures. Both are indicated in red. }\end{figure}

The first four terms constitute together the variational free energy of an \textit{annealed branched homopolymer} in the
Flory approximation \cite{Annealed_Branched, Disordered_UFN}. The branched homopolymer representation for vRNA molecules was
developed in ref. \cite{Fang} where the secondary structure of vRNA molecules was approximated as a collection of $N$
identical rigid segments of length $l$ connected by freely-jointed triple junctions into a tree-like structure. Analysis of
RNA secondary structures \cite{Fang} indicates that a reasonable choice for $l$ is about six nucleotides. For a 4,000 base
vRNA molecule, the number of segments $N$ is then in the range of $10^3$ (more details are provided in Supplemental Material, Sec. I [30]).
The different possible configurations of the branched polymer represents the different possible secondary structures.
Numerical evaluation of the enthalpy of vRNA molecules shows that there is a very large number of secondary structures with
enthalpy within $k_BT$ of the groundstate \cite{Fang}.
Two examples of tree-like structures with the same number of segments ($N$=21) -- but different MLDs (6, respectively, 11)- -
are
shown in Fig.\ref{Toy2}.

Returning to Eq.(\ref{fmix}), the first term is the entropic elastic free energy of a \textit{linear} homopolymer of $S$
segments with a radius of gyration $R$.  The second term is the conformational entropic free energy of an $N$-segment
branched polymer whose MLD equals $S$. The third and fourth terms represent the interactions between the segments expressed
in the form of a virial expansion in powers of the segment density $N/R^3$. The coefficient $V$ of the second-order term, which has the
dimension of volume, is typically of the order of $l^3$. It can be positive (``good solvent'') or negative (``bad solvent'').
The coefficient $W$ of the third-order term -- which represents the strength of three-body interactions - must be positive to
ensure thermodynamic stability. It is typically of the order of $l^6$. Minimization of the sum of the first four terms with
respect to $S$ and $R$ leads to a smooth coil-to-globule condensation transition around $V=0$. In the good solvent case,
the
radius of gyration scales with the number of segments $N$ as $R(N)\propto N^{7/13}$, which means that the swollen, or coil,
state has a fractal geometry. In the condensed phase, with negative $V$, the globule size scales as a compact object with
$R(N)\propto N^{1/3}$.  In either case, the MLD is determined  by the radius of gyration through $S(R) \sim
(N(R/l)^2)^{1/3}$. It follows that condensation decreases the MLD, thus increasing the amount of branching.

\subsection{Mixing entropy}

An important feature of the model is the fact that the second virial coefficient $V(x)$ of the branched polymer depends on
the probability $x$ that a segment is associated with a CP. In the model, precisely one segment can associate with the tail
of one CP so the maximum number of CPs that can associate with the branched polymer equals $N$. We will call such an $x=1$
aggregate a ``saturated aggregate''. Because CP-free vRNA molecules are known to be swollen under conditions of neutral pH
and
physiological salt concentrations, $V_0\equiv V(x=0)$ will be assumed to be positive.  On the other hand, in order for the
CPs to act as condensing agents for vRNA molecules, $V_1\equiv V(x=1)$ should be negative. The model assumes a linear
interpolation $V(x)=V_0-x(V_0-V_1)$ between these two limits (the bare second virial coefficient $V_0$ of a branched
flexible polyeletrolyte is roughly estimated as $l^3$. The second virial coefficient $V_1$ of a saturated aggregate, a more
complex quantity, is discussed and estimated in Supplemental Material Sect. II [30]).

Returning to Eq.(\ref{fmix}), the fifth term represents the binding affinity of a CP with a segment while the sixth term is
the
entropy of distributing $Nx$ different CPs over $N$ different segments.

\subsection{Electrostatic free energy}

The last term $ F_{PB}$ of the variational free energy is the electrostatic free energy of the aggregate as obtained from
Poisson-Boltzmann (PB) theory (see Supplemental Material, Sec. III [30] for a discussion of PB theory in the context of the model). The
macroion charge distribution is assumed to be as follows. The tail group is assigned a
charge $eZ$ and the head group a charge $-eZ$, with $Z\sim10$, while the segments of the branched polymer are assigned a
negative charge of $-eZ$, so one tail group can neutralize one segment. The total macroion charge of the branched
polyelectrolyte molecule equals $-NZ$ independent of the number of associated CPs. These macroions are placed in a
monovalent
salt solution with ion concentration $2c_s$. In PB theory, the electrostatic free energy of a macroion is determined by the
\textit{charging parameter} $\alpha$, which is defined as the ratio of the effective macroion charge $Q^*$ contained in a
certain volume $\Omega$ over the number $2c_s \Omega$ of monovalent salt ions in that same volume in the absence of the
macroion charge. The effective macroion charge differs from the bare charge because the monovalent salt ions can condense
onto the macroion and thereby diminish the effective charge. Within PB theory, the effective charge per unit length of a
highly charged polyelectrolyte molecule equals $-e/l_B$, where $l_B$ is the Bjerrum length defined by $ e^2/\varepsilon_0
l_B
= k_BT$ with $\varepsilon_0$ the dielectric constant of water. For the present case, the effective charge $Q^*$ of the
branched polymer equals $-(e/l_B)N/l$. For a vRNA molecule of 4,000 nucleotides confined to a sphere with a radius $R$ of
the
order of \unit[10]{nm}, the charging parameter $\alpha =\left| Q^{\ast} \right|  /2\Omega(R)c_s$, with $\Omega(R)=(4/3)\pi
R^3$, is of the
order of one.

The PB electrostatic free energy of capsid assembly has been extensively discussed (e.g., \cite{Siber1, Siber2}). In the
limits of small and large charging parameters it is given by (see Supplemental Material, Sect. III C [30])
\begin{equation}\label{clp4b4}
\begin{split}
&\beta{F_{\mathrm{PB}}}(R) \sim \frac{l^2}{\kappa} \frac{N^2}{R^3} \quad\quad\quad\quad\quad\quad\quad\alpha(R) \ll
1\\&\beta{F_{\mathrm{PB}}}(R) \sim2Q^{\ast} \ln \left( \frac{Q^{\ast}}{\Omega(R)c_s} \right) \quad\quad \alpha(R) \gg 1
\end{split}
\end{equation}
Here, $\kappa^2\equiv4\pi c_s e^2/(\varepsilon_0 k_BT)$ is the square of the Debye screening parameter. Note that, in the
weak-charging limit, $ F_{PB}(R)$ has the same form as the second virial term in Eq.(\ref{fmix}).
 \footnote{Under physiological conditions $1/\kappa$ is of the order of one nanometer.}

\subsection{Phase diagrams}

In order to obtain the phase diagram, $F(R,S,x)$ is first minimized with respect to $R$ and $S$ for fixed occupancy $x$. The
resulting free energy $F(x)$ has in general either one minimum or two minima separated by a maximum. For values of $x$ near
the maximum, the system is thermodynamically unstable and the solution decomposes into aggregates with different values of
$x$. As the magnitude $-V_1$ of the negative second virial coefficient is reduced, then the two minima of $F(x)$ approach
each other and merge at a critical value $V_c$. Define ${ \left< x \right> }$ to be the mean occupancy, i.e., the average of
the microscopic variable $x$ over all aggregates in solution. The mean occupancy is determined by the condition of phase
equilibrium between CPs that are associated with the branched polyelectrolyte and those that are free in solution. Equating
the chemical potential $\mu$ of the CPs in solution to the derivative $\frac{\partial F(x)}{N\partial x}$ of the free energy
of CP that is part of an aggregate with respect to the number $xN$ of CPs leads to a condition from which the mean occupancy
$\left< x \right>$ can be obtained by a common-tangent construction.

\subsubsection{Two types of disproportionation phase diagrams}
\begin{figure}[htb1]
\centering
\includegraphics[width=0.45\textwidth]{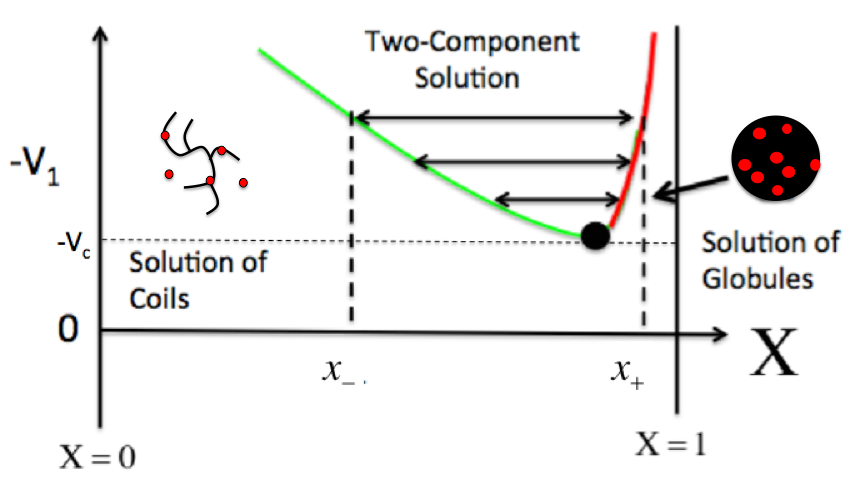}\\
\caption{(Color online) \label{Figure6} Disproportionation of a mixture of CP proteins and RNA molecules for large
binding affinities $\varepsilon$ (weak-charging regime). The horizontal axis is the CP to vRNA concentration ratio $X$.  The
vertical axis $-V_1$ is minus the effective second virial coefficient of a saturated globule. If $-V_1$ exceeds the critical
value $-V_c$ then phase decomposition takes place for mixing ratios in the interval $x_{-} < X < x_{+}$. The solid dot
indicates a critical point.}\end{figure}

 The shape of the resulting phase diagram depends crucially on the charging parameter. For the weak-charging regime and
 $\beta\varepsilon$ large compared to one, the phase diagram is shown in Fig.\ref{Figure6}. In this regime, the mean
 occupancy $\left< x \right>$ can be equated to the macroscopic CP to RNA concentration ratio $X=\phi_{CP}/N \phi_{RNA}$,
 normalized
 so $X=1$ corresponds to the concentration ratio of a saturated aggregate. We will assume that $X$ is less than or equal to
 one (as is the case for the experiments discussed in the conclusion).

The horizontal axis is the CP to vRNA concentration ratio and the vertical axis is the negative of the second virial
coefficient $V_1$ of a saturated aggregate. The solid dot indicates a critical point $V_1=V_c$ that marks the onset
of
the phase decomposition \footnote{$(\left| V_c \right| )^2/W\sim 1$.}. The interval of phase-decomposition widens as the
strength of the
negative second virial coefficient increases.

This phase diagram closely resembles that of the \textit{phase separation} of a polymer solution into dense and dilute
phases
when the solvent quality changes from good to bad \cite{RedBook}. There is however an important difference in terms of
interpretation. If the solvent quality is reduced in a polymer solution then the formation of globules typically induces
macroscopic phase separation, as is the case when condensing agents are added to a solution containing ssDNA molecules
\cite{Molas, Santai}. However, macroscopic phase separation does not (and should not !) occur during viral assembly. The
reason is that the
aggregates remain highly charged since the CPs, acting as the condensing agents, are charge neutral (at least in the model).
From this it follows that the CP-rich and CP-poor moieties in the two-phase region of the phase diagram remain mixed
together
in a dispersed state. Decomposition without macroscopic phase-separation is in fact well known from the literature on
complexation of oppositely charged polyelectrolytes as disproportionation \cite{Kabanov_disproportionation,
ZhangShklovskii_disproportionation} and we have adopted this usage.

\begin{figure}[htb1]
\centering
\includegraphics[width=0.4\textwidth]{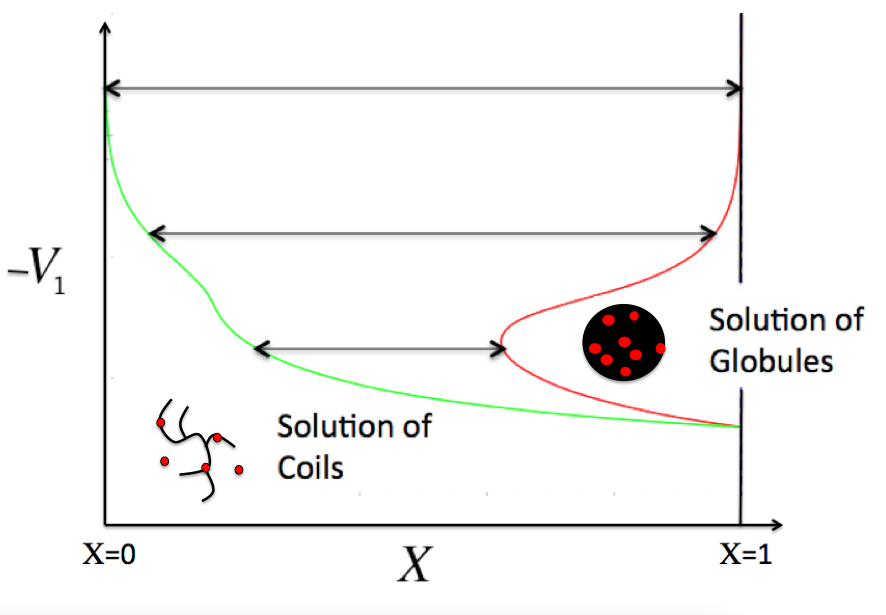}\\
\caption{(Color online) \label{fig:Segregation2} Same as Fig. \ref{Figure6} but now in the strong-charging
regime.}\end{figure}

The phase diagram in the strong-charging regime is shown in Fig. \ref{fig:Segregation2}.
The critical point has been replaced by a sharp, first-order coil-to-globule transition along the saturated globule line
$X=1$. For $X$ less than one, this transition broadens out into a wedge of phase decomposition, much like any phase
transition of a single-component material tends to broaden into a phase-coexistence interval when impurities are mixed in.
Note the surprising ``re-entrance'': if $V_1$ is increased starting from the coil phase for $X$ near one then
disproportionation appears, disappears, and then reappears.

\subsubsection{Coil-to-globule transition}
\begin{figure}[htb1]
\centering
\includegraphics[width=0.4\textwidth]{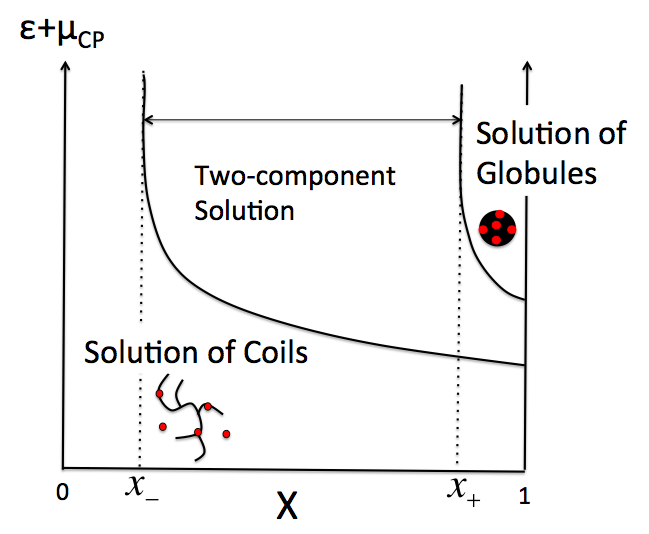}\\
\caption{(Color online) \label{Figure8} Phase-decomposition for fixed second virial coefficient $V_1$. The vertical
axis is the sum of the CP to RNA binding affinity $\varepsilon$ plus $\mu_{CP}$ where
$\beta\mu_{CP}=\ln(\phi_{\mathrm{CP}}/c_0)$. Here, $\phi_{CP}$ is the total CP concentration and $c_0$ the CP concentration
for a densely packed array of capsids. The horizontal axis is the CP to RNA mixing ratio $X$. The boundaries of
decomposition
$x_-$ and $x_+$ for large binding affinities are those shown in Fig. \ref{Figure6}}\end{figure}

As the binding affinity is reduced, the phase-diagram becomes dependent not just on the concentration ratio but also on the
total concentrations. This is shown in Fig.\ref{Figure8}, which displays the dependence of the width of the
two-phase region on $\varepsilon$ and the CP concentration $\phi_{CP}$ for the case that $-V_1$ is larger than the critical
value $-V_c$ (see Fig. \ref{Figure8}).
The vertical axis is the sum of the CP to RNA binding affinity $\varepsilon$ and $\mu_{CP}$ with
$\beta\mu_{CP}=\ln(\phi_{\mathrm{CP}}/c_0)$. Here, $\phi_{CP}$ is the total CP concentration and $c_0$ is the CP
concentration for a densely packed array of capsids. For $X\simeq1$ the aggregate is in the condensed globule state for
larger $\beta\varepsilon$.  When $\beta\varepsilon$ is reduced the solution decomposes into one moiety with aggregates whose
occupancy $x=x_+$ is close to one (the globule state) and one moiety whose occupancy $x_-$ close to zero (the coil state).
For sufficiently low $\beta\varepsilon$, the system is again in a one-phase region, but now with most CPs in solution and
with the vRNA molecules in the coil state. The coil-to-globule transition is smeared out as a function of
$\beta\mu_{CP}=\ln(\phi_{\mathrm{CP}}/c_0)$ because $\mu_{CP}$ is not the true CP chemical potential (see
Supplemental Material, Sect. V [30]). For large $\varepsilon/k_BT$, disproportionation is determined only by the
concentration ratio $X$, as we saw earlier.

\subsection{Surface Segregation}
\begin{figure}[htb1]
\centering
\includegraphics[width=0.45\textwidth]{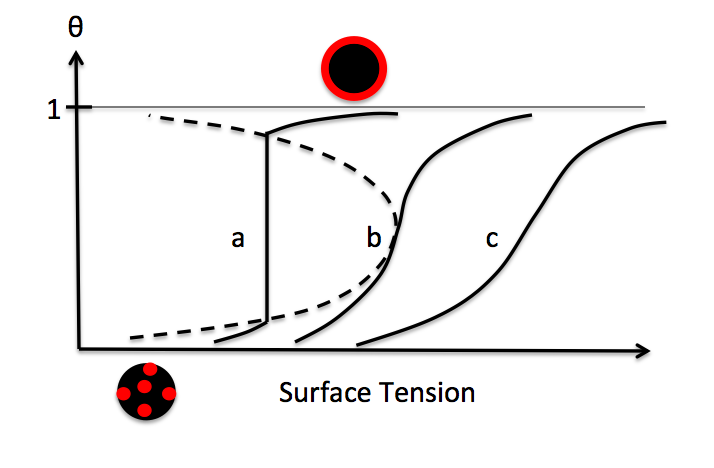}\\
\caption{(Color online) \label{fig-Segregation9} The fraction of CPs segregated to the surface of a saturated aggregate as a
function of the dimensionless surface tension $\beta\gamma_0 D^2$ of the globule. The curves a-c correspond to
decreasing attraction $u$ between oriented head groups located on the surface. Curve b corresponds to the critical isotherm.
}\end{figure}

As $-V_1$ increases, the system enters deeper into the condensed phase. A CP-RNA aggregate can no longer be treated as
uniform when this happens. The reason is that the head groups will segregate out to the surface of the condensate. There are
two reasons for this. First, condensed
globules have a \textit{surface tension} $\gamma_0$ \cite{RedBook} (a simple mean-field argument (see Supplemental
Material, Sect. VI [30])
gives $\beta\gamma_0 \sim{V}^2/W^{4/3}$). Head groups located in the interior effectively increase the
globule surface area. Transferring a head group from the interior to the surface lowers the free energy by an amount
$\gamma_0 D^2$, which we estimate to be of the order of $k_BT (D/l)^2$. Next, because head groups transferred to the surface
are oriented by the
surface - because the tail groups remain bound to the RNA interior - surface segregation also leads to a gain in
orientation-dependent attractive interaction between CPs. Figure \ref{fig-Segregation9} shows the fraction $\theta$ of CPs
located on the surface as a function of the dimensionless surface tension $\beta\gamma_0 D^2$ for different values of the
strength $u$ of the orientation-dependent attractive interactions between adjacent CPs located on the surface. The curves
were obtained using a Langmuir surface-adsorption model with attractive nearest-neighbor interaction for adsorbed particles
(see Supplemental Material, Sect. VIII [30] for details).
The curves marked $a-c$ are lines of fixed strength for the attractive interaction between surface-oriented CPs. Note
that the curves resemble the isotherms of the van der Waals gas. If $\theta$ is close to one, then the CP layer has the
character of a strongly correlated two-dimensional (2D) fluid while for $\theta$ close to zero it has the character of a weakly correlated 2D
gas. The transition between these regimes can be smooth (case $c$) or discontinuous (case $a$). A critical point located on the
$b$
isotherm separates the two regimes.

Surface-segregation leads to a \textit{geometrical conflict}. Let $R$ be the radius of a condensed globule with no head
groups in the interior. If the globule surface area $4\pi {R}^2$ is less than the area $ND^2$ of a layer of close-packed CP
head groups, then only a fraction of the CPs of a saturated aggregate can be accommodated on the surface. The surface of the
vRNA globule of $T=3$ ssRNA viruses accommodates about 180 CPs while a swollen CCMV vRNA molecule can accommodate about 300
CCMV CPs. One solution to resolve the conflict is for the excess CPs to be expelled into the surrounding solution at the
expense of losing an affinity $\varepsilon$ per tail group. This corresponds to the provirion 1 state of Sect. {I}. If
$\varepsilon$ is increased then breaking the bond between CPs and the vRNA molecule becomes too costly. Instead the surface
area of the condensate can be increased to allow access to the surface for more CPs. This corresponds to the provirion 2
state of Sect. {I}.  In the next section we extend the model to study the competition between the provirion 1 and 2
states,
assuming that surface segregation.

\section{Extended Model: Provirion States}
\label{EM}

The extended model is defined by separate free energies for the surface and the interior. The surface free energy area
density is defined as
\begin{equation}\label{987}
\begin{split}
\beta{{f}_s(\rho_2)}  = & \rho_2\ln\left(\frac{\rho_2 D^2}{1-{\rho_2 D^2}}\right)+B_{\psi}\rho_2^2+\\&+2\rho_2 Z
\ln\left(\frac{\rho_2 Zl_B}{\kappa}\right)-\beta\epsilon\rho_2
\  \end{split}
\end{equation}
The first two terms of Eq.(\ref{987}) constitute the van der Waals free energy density of a two-dimensional system of
disk-like
particles with area density $\rho_2$ and excluded area $D^2$. The last two terms are, respectively, the CP electrostatic
free
energy in the strong-charging limit (see Supplemental Material, Sect. III B [30]) and the CP-vRNA affinity. The (negative)
second
virial coefficient $B_{\psi}$ represents the hydrophobic pairing attraction between surface-segregated CP headgroups. It
depends on the angle $\psi=D/R$ between the relative orientations of the two axes of adjacent CPs (see Fig. \ref{fig:CCMV})
as
\begin{equation}\label{987D}
B_{\psi} /D^2 = -\exp [\beta u-(\psi-\psi_c)^2/\Delta\psi^2]
\end{equation}
Here, $u$ is -- as before -- the binding energy of the pairing attraction, $\Delta \psi$ is the angular range of the pairing
attraction, and $\psi_c=D/R_c$ is the relative angle between the CPs of a completed $T=3$ capsid with $R_c\simeq
\unit[10]{nm}$
the inner radius of the shell \footnote{the critical isotherm discussed at the end of the last section is determined by the
condition $e^{\beta u} \simeq Z$}.

The surface free energy $F_s=f_s(\rho_2) A$, with $A$ the globule surface area, must be added to the interior free energy
$F_b=f_b(\rho_3,x) \Omega$, with $\Omega\simeq (4/3)\pi {R_c}^3$ the globule volume, $\rho_3=N/\Omega$ the interior segment
density, and $x$ the segment occupation probability. The interior free energy density of a highly condense globule also has
the van der Waals form:
\begin{equation}
\label{Q}
\beta f_{b}(\rho_3, x) = \rho_3\ln\left(\frac{\rho_3/\rho_m}{1-\rho_3 /\rho_m)}\right)-a x\rho_3^2
\end{equation}
Here, $\rho_m$ is the maximum segment packing density, corresponding to a hydrated crystal of duplex RNA \footnote{In the
provirion states the vRNA density is so large that the virial expansion of Sect. II breaks down.}.  If we demand that the
radius of a close-packed sphere of vRNA segments equals \unit[9]{nm} then $\rho_m l^3\simeq 0.37$. The second term describes
tail group mediated attractive interaction between RNA segments where $x=\rho_2(A/N)$ is the occupation probability for a
vRNA segment to be occupied by a tail group. For provirion states with $\rho_2D^2\simeq1$, the occupancy $x\simeq A/ND^2$
reduces to a dimensionless measure of the surface area. In the limit of small $\rho_3$, the van der Waals free energy
reduces
to our earlier virial expansion with a second virial coefficient $V(x)=(1/\rho_m) - a x$ that decreases linearly with the
occupation probability \footnote{$W=(1/\rho_m)^2$}. As in Section II, the second virial coefficient will be assumed to
change
sign as a function of $x$, separating states where the aggregate is in good solvent (for smaller $x$) or in bad solvent (for
larger $x$). As before, the globule is assumed to have a surface tension $\beta\gamma_0(x) \sim{V(x)}^2/W^{4/3}$. However,
the second virial coefficient $V(x)$ for the surface segregated state in general will be different from that of the uniform
globule state.

\subsection{CP-exchange equilibrium and surface phase diagram}
The next step is to impose thermodynamic equilibrium of the globule surface and interior, both with respect to each other
and
with respect to the surrounding solution. The head group surface area density $\rho_2$ is determined by the condition of
exchange or phase equilibrium between CPs located on the globule surface and in the surrounding solution. For simplicity, we
will assume in this section that the solution CP chemical potential $\mu$ is a fixed quantity. Exchange equilibrium is
satisfied if
\begin{equation}
\label{EE}
\beta\partial f_s/\partial \rho_2-a{\rho_3}=\beta\mu
\end{equation}
The second term on the left hand side is due to the fact that the interior free energy density, through $x$, also depends on
the surface CP density.
A surface phase diagram can be obtained in terms of the 2D surface pressure $\Pi_2=\rho_2\frac{{\partial
f}(\rho_2)}{\partial\rho_2}-{f}(\rho_2)$ and the strength $\exp-\beta u$ of the attractive interactions.
\begin{figure}[htb1]
\label{Figure9}
\centering
\includegraphics[width=0.4\textwidth]{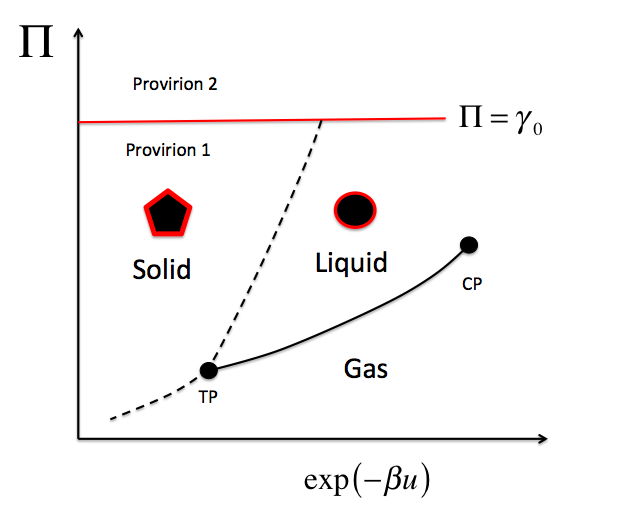}\\
\caption{(Color online) \label{Figure9} Phase diagram of the surface layer. (Solid line) Liquid to gas transition, ending at
a
critical point (CP). (Dashed lines) Solidification and sublimation lines ending at the triple point (TP, not included in the
model). The thermodynamic surface tension vanishes along the red line.}\end{figure}
The surface phase diagram has a line of first-order transitions separating a liquid and a gas phase, as shown in Fig.
\ref{Figure9}, ending at a critical point (CP) where $\exp\beta u\simeq Z$. In a more complete model, this phase diagram
would also contain other phases with, minimally, a solidification line ending at a triple point (TP), where it joins a
sublimation line (both marked as dashed lines). The horizontal red line will be discussed below. We will restrict ourselves
to the high-density liquid phase with $\rho_2 D^2\simeq 1$ and $\rho_3/\rho_m\simeq 1$. The solution for Eq.(\ref{EE})
corresponding to those condition has a surface pressure $\Pi_2\sim (\mu+\epsilon)/D^2$ that increases (approximately)
linearly with $(\mu+\epsilon)$ \footnote{ $\beta \Pi_2 D^2\simeq\beta(\mu+\epsilon)+\exp(\beta u)-2Z
\ln \left( \frac{Zl_B}{D^2\kappa} \right) + a {\rho_m}-\ln \left( \beta ( \mu + \epsilon) + \exp(\beta u)-2Z
\ln \left( \frac{Zl_B}{D^2\kappa} \right) + a{\rho_m} \right)$} .

\subsection{Mechanical equilibrium}
The next step is to impose \textit{mechanical equilibrium} which requires that the total free energy is minimized with
respect to $\rho_3$. This leads to
\begin{equation}\label{987E}
\frac{{2(\gamma_0(x)-\Pi_2})}{R}-\frac{K_H}{R_c} \left[ \frac{2}{R} - \frac{2}{R_c} \right]^3=\Pi_3(\rho_3)
\end{equation}
Here, $\Pi_3=-\partial F_b/\partial \Omega$ is the three-dimensional (3D) osmotic pressure exerted by the interior on the surface layer. The
first term on the left hand side can be understood by noting that $\gamma =
\gamma_0-\Pi(\rho_2)$ is the \textit{thermodynamic surface tension} -- defined as $\gamma=\partial F/\partial A$ -- so
$2\gamma/R$ can be interpreted as a \textit{Laplace pressure}. Under conditions of thermodynamic equilibrium, the
thermodynamic surface tension is related to the chemical potential by the Gibbs isotherm $d\gamma = -\rho_2 d\mu$ with $\mu$
again the chemical potential.

The second term in Eq.(\ref{987E}) is a pressure that is generated by the dependence of the second virial coefficient
$B_{\psi}$ on angle, and hence on $R$. The same term would have been obtained if we had included a \textit{Helfrich bending
energy} in the surface energy with mean curvature $2/R$ and spontaneous curvature $2/R_c$ (see also ref.\cite{Siber3}).
Here,
$\beta K_H={\exp(\beta u)}{/\Delta\psi^2}$ acts as a dimensionless bending modulus. Equation (\ref{987E}) can be extended to
non-spherical surfaces, by replacing $2/R$ with the mean curvature. Finally, the pressure $\Pi_3(\rho_3)$ in Eq.(\ref{987E})
exerted by the interior on the surface is given by the usual van der Waals equation of state:
\begin{equation}\label{9876}
\beta\Pi_3(\rho_3)=\frac{\rho_3}{1-\rho_3/\rho_m}-ax{\rho_3}^2
\end{equation}
We will only consider solutions of Eq.(\ref{987E}) with $\rho_2 D^2\simeq 1$ and $\rho_3/\rho_m\simeq 1$. The nature of the
solution depends in this case on which term dominates the left hand side of Eq.(\ref{987E}).

\subsubsection{Provirion 1 state}
First assume smaller values for $\epsilon+\mu$ and larger values for $u$ so the second term, the Helfrich pressure term,
dominates over the Laplace pressure term. The dominant Helfrich energy is minimized if the shell adopts the geometry of a
sphere with radius equal to the spontaneous curvature radius (here $R_c$). The segment occupation probability $x$ can be
equated to $x_v=4\pi R_c^2/N D^2$, the occupation probability of an assembled virion (for CCMV, $x_v$ is of the order of
$0.6$). A significant fraction of the RNA segments are not associated with a tail group in this state, which reflects the
geometrical
conflict we noted earler. The second virial coefficient has increased from $V(x=1)$ to $V(x=x_v)$ when $x$ is reduced from a
value close to one to $x_v$. Similarly, the bare surface tension of the globule must be reduced, say to $\gamma_0(x_v)$ If
$V(x_v)$ still is negative,
like $V(x=1)$, then the RNA material remains in the condensed state. The interior osmotic pressure $\Pi_3$ exerted on the
shell can be neglected in this case. The thermodynamic surface tension may be positive or negative, depending on the sign of
$\gamma_0(x_v)-\Pi_2(\epsilon+\mu)$.  Surfaces with a negative surface tension normally are thermodynamically unstable but
the Helfrich bending energy can suppress this instability for sufficiently large bending energy.
If, on the other hand, the second virial coefficient $V(x_v)$ of the interior is positive then the interior is in a good
solvent state and exerts
a positive osmotic pressure on the shell. The bare surface tension is zero in this case. The combined pressures
$\Pi_3+2\Pi_2(\epsilon+\mu)/R_c$ must be adsorbed by the Helfric bending energy.

\subsubsection{Provirion 2 state}
For increasing $\mu+\epsilon$ the surface pressure $\Pi_2$ rises. When $2\Pi_2/R_c$ approaches $K_H/R_c^4$ in magnitude then
the Helfrich bending energy is no longer able to compensate for the surface pressure. The surface area is forced to expand
until the occupation probability reaches its maximum value $x=1$. The interior remains condensed since $x=1$ so the interior
pressure $\Pi_3$ can be set to zero. Because the surface area $A\simeq ND^2$ now exceeds $4\pi (R_c)^2$ the surface cannot
remain spherical. By analogy with similar problems in the physics of surfactants \cite{Safran}, we expect that the mean
curvature will remain close to $2/R_c$ over sections of the surface that are bounded by lines of negative Gauss curvature where
that is not the case.

\section{Summary and Conclusion}

In this concluding section, we first compare a number of predictions of the model with the outcome of recent experiments on
self-assembly of the CCMV virus. We then discuss predictions of the model that have not yet been tested and conclude with
the
most important limitations of the model.

\subsection{Comparison with experiment}

The most distinctive prediction of the model in terms of experimental tests concerns the \textit{optimal mixing ratio}
(OMR),
defined as the minimum value of the CP-to-vRNA concentration ratio $X$ for which all of the vRNA molecules are packaged. The
OMR has been measured through virion assembly experiments in solutions that contained CCMV CPs and non-CCMV vRNA molecules,
using an assembly protocol aimed at maintaining thermodynamic equilibrium \cite{Cadena-Nava_2012}. The non-native vRNA
molecules had the same length as that of CCMV vRNA molecules. Solutions with a prescribed mixing ratio were first incubated
at neutral pH and low salinity, so with weak CP-CP pairing attraction. Cryo-EM images of the solution revealed the formation
of virion-sized complexes of CP and RNA with irregular and disordered shapes. RNase digestion assays showed that these
disordered complexes did \textit{not} protect the RNA from degradation by RNase so the complexes could not be stable
virions.
CP-RNA binding was reversible and CPs could exchange between different RNAs \cite{Comas-Garcia}. When CP-CP interactions
were
strengthened by lowering of the pH from 7.2 to 4.5, true virions formed from these structures.

\begin{figure}[htb]
\centering
\includegraphics[width=0.30\textwidth]{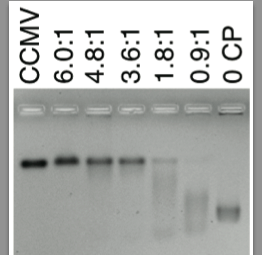}\\
\caption{\label{fig:RNACCMV} CP-RNA assembly titrations: gel retardation assays.  Shown are 1$\%$ agarose gels run at low pH
and stained for RNA.  At the left is a titration of 3217nt RNA1 molecules of the Brome Mosaic Virus (BMV) with varying
amounts of CCMV CP. The value of the CP to RNA weight ratio $w$ is provided at the top of each lane. It ranges from 0
(right-most lane, RNA) to 6:1 (lane second from left). The weight ratio $w$ is related to the mixing ratio $X$ of the text
by
$w\simeq6.0 X$. The left-most lane shows the position of  CCMV virions. From ref.\cite{Cadena-Nava_2012}}\end{figure}

The results of
electrophoresis
runs for different mixing ratios $X$ \cite{Comas-Garcia} are shown in Fig.\ref{fig:RNACCMV}.
The far left column shows the case of native CCMV. The far right column shows the case of solutions containing only RNA
molecules, which thus move faster than native CCMV virions during gel electrophoresis. For the case of CP to RNA weight
ratios $w$ below $6.0$, a narrow band moves with a velocity slightly less than that of CCMV virions. The weight rato $w$ is
related to the mixing ratio $X$ of the previous sections by $w\simeq6.0X$ so $w\simeq6.0$ corresponds to $X\simeq1.0$. This
suggests that aggregates in this band at least resemble the CCMV virion. These labile aggregates could not correspond to the
provirion 2 state, since the provirion 2 state is larger than the CCMV virion state, and are candidates for the provirion 1
state.

The faster RNA band has broadened out extending from velocities higher than that of the pure RNA molecule, down to the
CCMV-like band. Aggregates in this smeared out band are not packaged when the interaction strength is increased, so $X$ is
less than the OMR. If the CP concentration is increased then the broad band disappears around a concentration ratio of about
$300$ CPs per vRNA molecule. For a positive tail charge of about $+10e$, this corresponds to an OMR of $X=1$.

This is a striking result. If one would apply textbook self-assembly theory \cite{Safran} then -- by directly minimizing the
free energy of a solution of CPs and vRNA molecules in the absence of any CP-induced vRNA condensation -- one obtains
Fig.\ref{Assembly} (see Supplemental Material, Sect. IV [30]).
\begin{figure}[htb]
\centering
\includegraphics[width=0.30\textwidth]{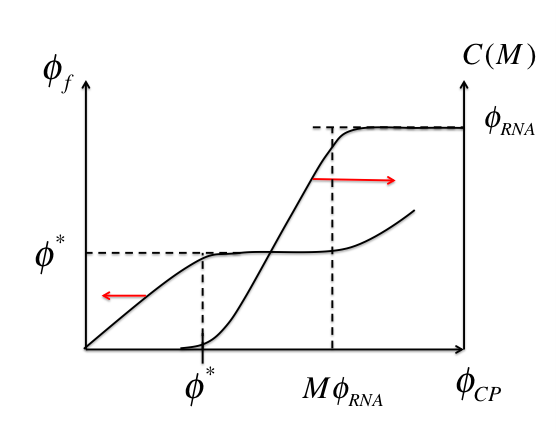}\\
\caption{(Color online) \label{Assembly} Dependence of the concentration $\phi_f$ of free capsid proteins and that of the
capsid concentration ${C}(M)$ on the total protein concentration ${\phi}_{\mathrm{CP}}$ (left vertical axis) according to
textbook theory. Capsid assembly starts at the CMC $\phi^{\ast}$ and terminates when the RNA supply has been
exhausted at ${\phi}_{\mathrm{CP}}\simeq M{\phi}_{\mathrm{RNA}}$ with $M$ the number of CPs per virion.}
\end{figure}
As a function of increasing CP concentration, capsid assembly starts at a CMC, denoted by $\phi^{\ast}$, which is
proportional
to the Boltzmann factor for inserting a CP into a virion shell. Beyond $\phi^{\ast}$, the concentration of free CPs
saturates
while that of asssembled capsids increases linearly with the CP concentration. The increase stops when the supply of vRNA
molecules is exhausted, which is the OMR at which (nearly) all vRNA molecules have been packaged. The OMR is thus
$X=M/N$, with $M$ the number of CPs of a $T=3$ shell and $N$ the number of vRNA segments (see Supplemental Material, Sect. VII [30] for a more detailed discussion).

The value of the OMR predicted by the model follows from Figs. 6-8. These show that the solution
should disproportionate into CP-rich globules and CP-poor vRNA molecules. Provided the CP-rich globules are in the provirion
1 state, the globules should  transform into virions when the strength $u$ of the CP-CP pairing is increased and the 2D
liquid freezes into a $T=3$ ``crystal''. The CP-poor swollen vRNA molecules will not be packaged. It follows that, for the
model, the OMR is $X=1$. More generally, the OMR corresponds to charge neutralization of the vRNA molecule by the CP tail
groups. Measurement of the OMR is thus a direct way compare the theory proposed in this paper for virion assembly and
textbook self-assembly theory.

If one interprets the smeared-out band in Fig.\ref{fig:RNACCMV} as being produced by CP-poor swollen vRNA globules then
there
would seem to be agreeement between the predictions of the proposed model and experiment and a direct violation of
conventional self-assembly theory. It should be recalled here that conventional self-assembly theory works quite well for
the
assembly of empty capsids. This interpretation can (and should) be questioned. In a solution that is in \textit{complete}
thermodynamic equilibrium, each vRNA molecule should fluctuate thermally between all of the allowed configurations. In an
electrophoresis experiment that is carried out on a system in full equilibrium, the vRNA molecules should \textit{all} move
with an average speed determined by a Boltzmann average over all accessible states, leading to just one single band. Figure
\ref{fig:RNACCMV} indicates that the previous experiments were carried out on time scales shorter than the thermal
equilibration time. Now, it seems reasonable to assume that the life-time of an assembled provirion state is the longest
relaxation time of the system. In an electrophoresis experiment carried out on time scales shorter than this relaxation time
but longer than any other relaxation time, one should expect a \textit{bimodal distribution} with two narrow bands. The slow
band contains provirions and the fast band corresponds to aggregates that interconvert among each other on the time scale of
the experiment. A recent assembly study of CCMV with much shorter 500-nt-long RNA fragments reported bimodal distributions
for nearly all CP:RNA ratios \cite{Garmann5}. The natural interpretation is that for the case of shorter RNA chains the
system is closer to thermal equilibrium. Recently, Kler and co-workers found -- for the SV40 virus -- that titration of a
short
RNA molecule (less than 0.8 kb) with VP1 indeed gave a bimodal distribution while binding of VP1 to longer RNAs again led to
the formation of intermediate species \cite{Kler_2013}. Bimodal distributions have also been observed in the
\textit{in-vitro} assembly of cucumber mosaic virus (CMV) \cite{Kobayashi_1995}. If the size of the vRNA molecule is
increased, then provirions that are missing variable amounts of CPs are expected to have relatively long life times. These
would show as a smearing of the slow band.

A second distinctive prediction of the model concerns the claim that the CP-tail groups must be effective vRNA condensing
agents. If the compression of the vRNA molecule was purely due to the action of the CP head groups -- which we argued against
in the introduction -- then removal of the CP head groups should cause \textit{swelling} of the vRNA molecule, while the
model
predicts increased condensation. In ref. \cite{McPherson, Yuri} it was shown that the vRNA molecules of the $T=1$ Satellite
Tobcacco Mosaic Virus (STMV) remained in a fully condensed state after the CP head groups had been enzymatically removed
from
STMV virions while the tail groups of the CPs remained behind. The resulting particles were thermodynamically very stable.
X-ray diffraction studies revealed a tight association between the tail groups and the vRNA molecule \cite{McPherson, Yuri}.

\begin{figure}[htb]
\centering
\includegraphics[width=0.45\textwidth]{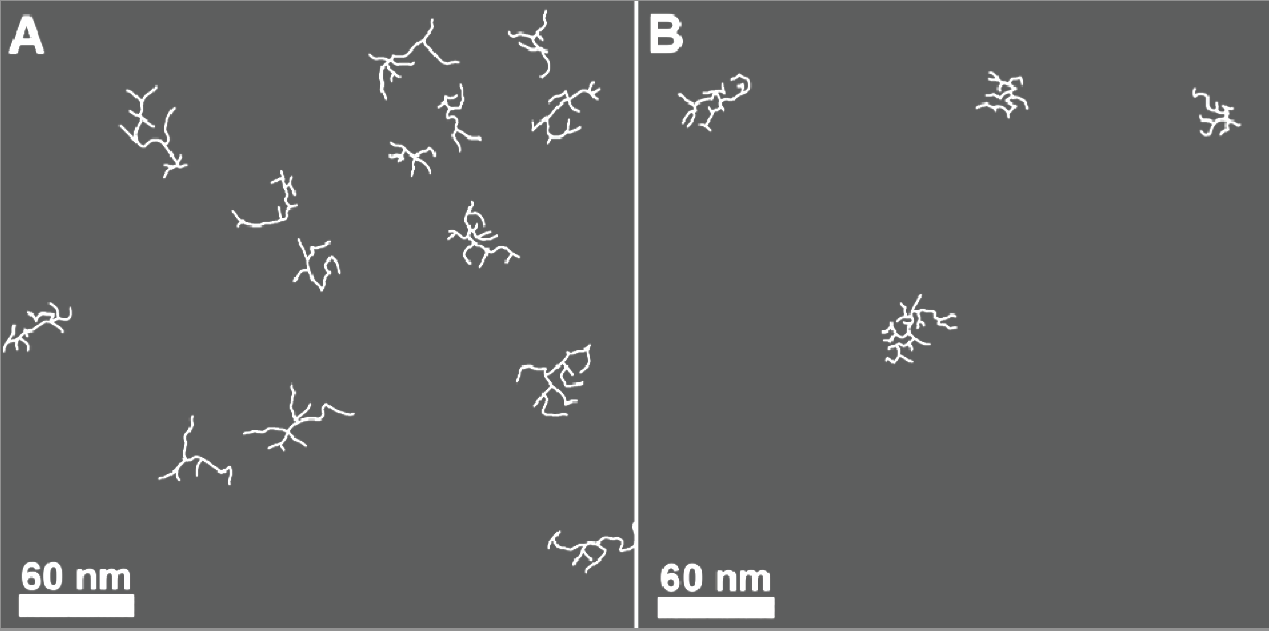}\\
\caption{(Color online) \label{cond} Cryo-EM images of 2777-nt RNA molecule under assembly conditions (left panel) and
assembly conditions with added $\mathrm{Mg}^{2+}$ ions.  The image is reproduced from ref.\cite{Gopal} with
permission}\end{figure}

Next, in the weak-charging limit, vRNA molecules in good solvent should have a fractal structure with a radius of gyration
$R(N)$ that scales with the number of monomers as $N^{1/2}$ while in the strong-charging limit, the molecules should be more
linear and extended, and the radius of gyration should be proportional to $N$. In the condensed state, the radius of
gyration
should scale as $N^{1/3}$. In either case, the MLD of the vRNA molecule should depend on the radius as $R^{2/3}$, which can
be tested experimentally. Gopal and coworkers visualized CCMV RNA 2 (of \unit[2.7]{kb}) molecules using cryo-electron
microscopy \cite{Gopal}. They found that, in a physiological buffer without $\mathrm{Mg}^{2+}$, the RNA molecules adopted
highly extended structures with just a few major branches. An example is shown in the left panel of Fig~\ref{cond}.
The appearance of extended vRNA structures suggests the strong-charging regime where the vRNA molecules are effectively
stretched by electrostatic repulsion. When the solution concentration of $\mathrm{Mg}^{++}$ ions was increased, more
compact,
spherical shapes appeared with a smaller radius comparable to that of the virus itself as shown in Fig.~\ref{cond}, right
panel. These molecules had structures that were significantly more branched than the swollen structures in good solvent.
These results seem at least consistent with the simple model of Sect. II.

\begin{figure}[htb]
\centering
\includegraphics[width=0.45\textwidth]{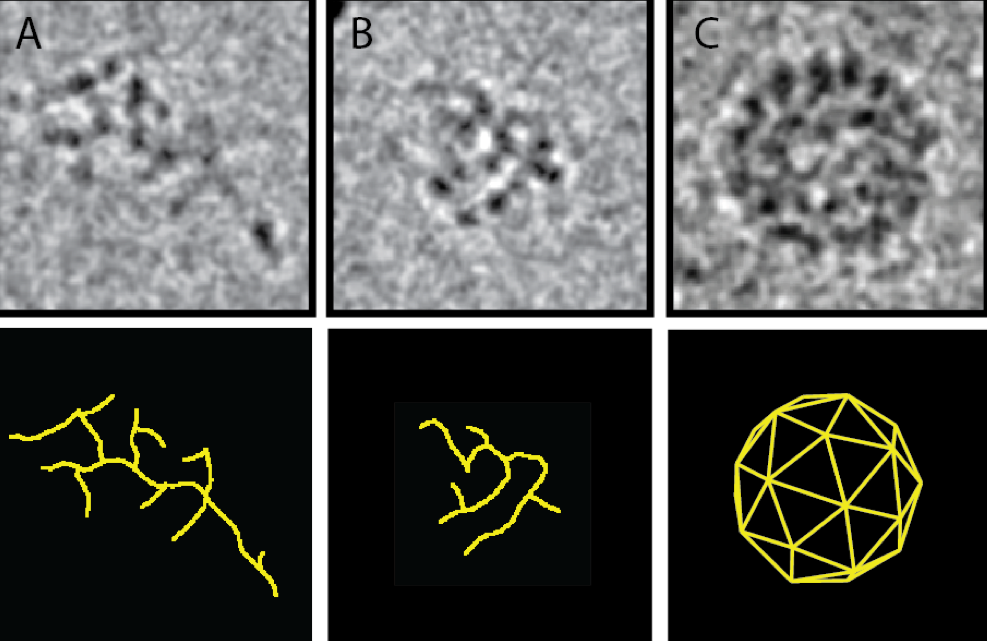}\\
\caption{(Color online) \label{provir} (Top) Cryo-EM images of 3,200-nt RNA during the different stages of assembly. Bottom:
reconstructions. (a) Shows naked RNA in assembly buffer with $\mathrm{Mg^{2+}}$ (see also Fig. \ref{cond}). (b) Shows the same RNA molecule but decorated with a super-stoichiometric amount of CP at higher pH. Note that the complex is
smaller than the naked RNA molecule. Analysis of a large number of cryo-EM images shows that the average size drops from
about \unit[37]{nm} when the RNA is naked to about \unit[32]{nm} in the provirion state when it is decorated by CP. (c) Shows the formation of capsid-like structures when CP-CP interactions are strengthened by reducing the pH.  Reproduced
from ref.\cite{Garmann} with permission from Elsevier.}\end{figure}

Has the provirion state been observed microscopically? Figure \ref{provir} shows cryo-EM images of 3,200-nt RNA molecules
when CPs are added to the solution.
Figure \ref{provir} (a) shows the RNA molecule in assembly buffer with added $\mathrm{Mg}^{2+}$ in the absence of CPs, as in the previous
picture. Note again the elongated arms, indicative of strong electrostatic repulsion. Figure \ref{provir} (b) shows the same RNA molecule
when the CP to RNA mixing ratio $X$ is larger than $0.6$. The images were taken at higher pH when the attractive
interactions
are too weak to support virion assembly. The approximated structure of the RNA molecule is shown in the bottom image. It
clearly has undergone a certain degree of additional condensation. The image shows -- probably transient -- \textit{shell
fragments}. If this would be the image of a provirion 1 then the description of the surface-segregated CPs as a 2D
correlated
fluid will have to be replaced by a more complex fluctuating state with a statistical distribution over capsid fragments of
various size. Figure \ref{provir} (c) shows the structure of the aggregate when the pH is reduced so the strength of the CP-CP attraction
increased. If \textit{this} is an image of a provirion 1 then it would correspond more closely to the description proposed
in
this paper, although it could also be already a true virion. A key point would be to determine at what pH the shell transforms
from a fluid state that is in thermodynamic equilibrium with the surrounding solution to an ordered $T=3$ capsid with
frozen-in
CP positions.

\subsection{Tests of the model}

We now turn to the predictions of the model that allow future experimental tests of its validity. The existence of a
provirion 2 state plays a central role in this respect. According to the general phase diagram (see Fig.3), increasing the
binding affinity $\varepsilon$ should lead to a stabilization of provirion 2 state with respect to the provirion 1 state.
Increasing $\varepsilon$ could be done by systematically increasing the number of positively charged arginine residues on the
CP tail groups, as was already done in ref.\cite{Garmann_2014}. Electron microscopy images of a provirion 2 state should be
characterized by strongly
fluctuating, non-spherical shapes. Increasing the strength of the CP-CP attraction starting from a provirion 2 state should
produce not virions but malformed structures, as in Fig. \ref{Toy2}.

Next, reducing $\varepsilon$ - for example by reducing the number of charged residues per tail group - would allow a second
experimental test of the proposed model. According to the assembly diagram, for smaller values of $\beta\varepsilon$,
assembly as a function of increasing $u$ should proceed without vRNA condensation.  According to Fig.\ref{Assembly}, the OMR
should then be $X=M/L\simeq 0.6$ (for the case of CCMV at least). Moreover, the fraction of assembled virions measured, as a
function of the total CP concentration, should now obey the Law of Mass Action, as is the case for empty capsids but as is
not the case in the model if assembly starts from a provirion precursor state.

Another important prediction of the model concerns the presence of either a critical point or a first-order phase transition
point in the assembly diagrams (see Figs. 8 and 9), depending on the charging parameter $\alpha$. This could be tested by
repeating the electrophoresis experiments discused above but now decreasing the magnitude of the negative second virial
coefficient $-V_1$. The second virial coeffcient of the CPs could be quantitatively measured separately by thermodynamic
studies of CP pair formation in dilute solutions of CPs. Variation of $V_1$ as a function of pH, salinity, or tail length
could then be determined. Measurement of the disproportionation interval -- in terms of the mixing ratio $X$ -- by gel
electrophoresis for different values of $V_1$ could verify whether the weak or strong-charging regime applied. Recall here
the striking prediction of reentrance of the single-phase region in the assembly diagram as a function of $-V_1$ for the
strong-charging case.

\begin{figure}[htb]
\centering
\includegraphics[width=0.25\textwidth]{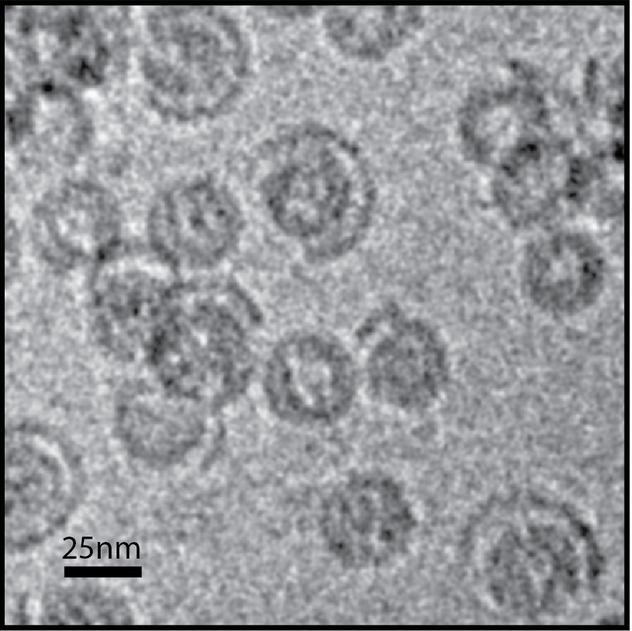}\\
\caption{(Color online) \label{double} Cryo-EM images of assembly products of 500-nt RNA. Multishell structures formed with
an inner shell that roughly correspond to a $T=2$ shell and an outer shell that corresponds to an incomplete $T=3$ shell.
Reprinted with permission from \cite{Garmann5}.  Copyright 2014 American Chemical Society.}\end{figure}

The experiments discussed above could be repeated for different salt concentrations. Reducing the salinity means increasing
the strength of the electrostatic interactions. Studies of the assembly of empty CCMV capsids \cite{Bancroft_1968,
Adolph-Butler, Lavelle} reported that at higher CP concentration and lower ionic strength, \textit{multishell structures}
form, stabilized by electrostatic interactions \cite{Prinsen_2010}, where the tail groups of the second layer associate with
the head groups of the CP first layer. It has been shown that multi-layer shell structures form during assembly of virions
with shorter RNA molecules, as shown in Fig. \ref{double} \cite{Garmann5}.
If the gel-electrophoresis experiments discussed above were repeated at lower salinity, then the excess CPs released in
solution would now be expected to remain associated with the CP shell in the form of a second layer (see Figs.4 and 5). By
measuring the number of CPs that remain associated with the virion after assembly - for example by fluorescent labeling - it
could be checked if the number of excess CPs equals the difference between the number of CPs of a saturated aggregate and of
a virion.

\subsection{Overcharging}
In the introduction we posed the question why the total positive charge of the CP tail groups of the CPs is not neutralized by the negative charge of the RNA molecule. Recall that overcharging in the context of viral assembly previously has been attributed to the local structure of tail group/RNA association in ref. \cite{Boris(2)} and to Manning condensation in ref. \cite{Belyi}. In the proposed model, overcharging is attributed to correlations between the CP head groups: The homogeneous saturated aggregate, in which the tail groups do neutralize the RNA charge, is ``frustrated'' by the non-electrostatic, angle-dependent interactions that drive the formation of the provirion 1 state and the $T=3$ capsid.

Could experiment resolve questions about the cause of the overcharging?  It is possible to enzymatically digest the head groups of a virion \cite{Day}. If the proposed explanations of either Refs. \cite{Belyi} or \cite{Boris(2)} are right, then the macroion overcharge of the remaining RNA/tail-group core particle should remain stable in a solution that contains a modest concentration of tail-group molecules, as it represents a minimum free energy state. On the other hand, in the model proposed here, the core particle should be expected to ``soak-up'' extra tail-groups from the surrounding solution causing the overcharge to decrease to zero. Changes of the charge of a core particle could be monitored in an electrophoresis experiment.

This method could also be used to measure the osmotic pressure the RNA/tail group assembly exerts on the capsid. If this pressure is negative, then the RNA/tail-group core particle ought to contract after enzymatic digestion of the capsid while in the opposite case, it should expand. The radius of the core particles in solution could be measured by AFM, as was already done \cite{Day}, or in a scattering experiment. By adjusting the osmotic pressure of the surrounding solution untill the radius of the core particle would equal the inner radius $R_c$ of the capsid, one could establish the osmotic pressure inside the virion. In the proposed model, the core particle should expand after digestion of the capsid.

\subsection{Limitations of the model}

We finish by discussing the limitations of the proposed model. The applicability of the proposed model to the complexity of
viral assembly involves both straightforward simplifications that could impair quantitative predictions but which can be
improved upon in a systematic fashion, and more ``dangerous'' assumptions whose failure would compromise the usefulness of
the
model at a fundamental level.

Straightforward simplifications involve equating the magnitudes of the head group and tail group charges, assuming a linear
dependence of the second virial coefficient on occupancy, assuming a Gaussian dependence of the surface second virial
coefficient on angle and assuming that the third virial coefficient of a CP/vRNA saturated aggregate does not depend on
occupancy. Higher-order terms may have to be systematically included in the virial expansion. In order to carry out
quantitative tests, these assumptions may have to be improved upon. However, we believe -- although have not explicitly
demonstrated -- that none of the key predictions discussed earlier will be affected if the model is generalized.

A more serious limitation of the model concerns the use of mean-field theory. We made the assumption that the interior of a
surface-segregated globule is homogeneous. In actuality, because CP tail-groups are attached to the surface-segregated CP
head-groups, neutralization of the negative vRNA charges must be more efficient near the surface of the globule than in the
interior. As a result, the macro-ion charge density will have a radial profile. Mean-field theories allowing spatial
variation of the density can be formulated but it would seriously complicate the formalism. Fluctuations around mean-field
theory, even one that includes a non-trivial density profile, are neglected as well. As discussed in the conclusion, if the
surface of a provirion 1s better described as a collection of transient shell fragments instead of a correlated but uniform
fluid then this would be a serious concern for the theory that would not be easy to remedy.

Another ``dangerous'' limitation of the model is the use of PB theory to describe the electrostatics. It can be shown that the
condensation of ds DNA molecules by polyvalent counterions is due to correlation attraction. This effect is beyond PB theory
\cite{Niels} and it is possible -- even likely -- that the same is true for the condensation of ss RNA molecules. This problem
was to some extent ``swept under the rug'' by including correlation attraction effects as negative contributions to the
effective second virial coefficient. The applicability of PB theory can be monitored by measuring the OMR. Serious breakdown
of PB theory would be signaled by the appearance of overcharging \cite{Nguyen_overcharging} of saturated aggregates since these would no longer correspond to a state in which the tail groups neutralize the RNA molecules. That would mean that $X=1$
would
not correspond to the OMR. For CCMV at least, that appears not to be the case but it could well be true for other viruses.

A final important limitation of the model is the restriction to equilibrium thermodynamics. The assembly of empty capsids
follows the law of mass action of equilibrium statistical mechanics. In actuality, empty capsids in fact do
not disassemble when the CP concentration in solution is reduced back to zero. The final assembly step or steps are
quasi-irreversible. This must be true as well for virions or else virions would disassemble in CP free solutions, which is
not the case. Kinetic models of empty capsid assembly confirm that a form of the law of mass action survives when only a few
number of steps are irreversible \cite{Morozov}. In general, equilibrium models are expected to fail progressively as the
number of irreversible assembly steps increases. It is our belief however that an understanding viral assembly in general
requires understanding viral assembly under conditions of thermal equilibrium.

\begin{acknowledgments}
We wish to thank Chuck Knobler for numerous suggestions, discussions and readings of the manuscript. We also wish to thank
Vinny Manoharan for a critical reading of the draft. We are grateful to Jack Johnson for providing the image used for Fig.
\ref{fig:CCMV(1)}.  We would like to thank as well Michael Hagan and Boris Shklovskii for discussions and comments. RB
thanks
the National Science Foundation (USA) for support under Grant No. DMR-1309423.  The work of AYG on this project was supported in part by the National Science
Foundation (USA) under Grant No. PHY-1066293. Both AYG and RB wish to thank the Aspen Center for Physics for its hospitality and
for hosting a workshop on the physics of viral assembly where some of the work was done.
\end{acknowledgments}

\bibliography{Virus_References_3}

\newpage

\section*{Supplementary material}

\section{Secondary Structure of Viral RNA Molecules}
\begin{figure}[htb]
\centering
\includegraphics[width=0.40\textwidth]{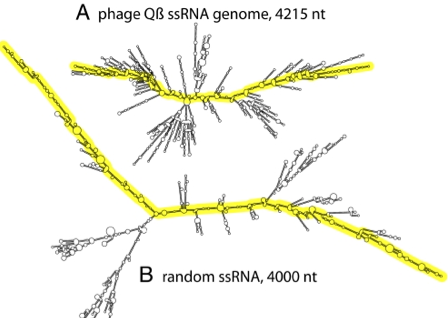}\\
\caption{\label{fig:Sec} Secondary structures of a vRNA genome molecule (A) and of an RNA molecule with a random sequence of nucleotides (B) with the same length. The longest end-to-end lengths $S$ of the molecules, known as the maximum ladder distance, are indicated by yellow lines.  Reprinted with permission from Ref. \cite{Aaron} (Copyright (2008) National Academy of Sciences, U.S.A.).}\end{figure}

The secondary structure of single-stranded RNA molecules is determined by the complementary pairing of nucleotides. Figure~\ref{fig:Sec} shows a typical minimum free energy secondary structure of a T=3 viral RNA molecule obtained by applying the Mfold program~\cite{Aaron}.
The top figure is that of a vRNA molecule of about 4,000 nucleotides. The bottom figure is that of an RNA molecule that has the same number of nucleotides but with a randomly chosen nucleotide sequence. Both molecules are composed of short complementary paired sequences that alternate with unpaired bubbles and branch points. The size of the molecule is determined by the \textit{maximum ladder distance} $S$, defined as the longest distance between ends of the branched structure, counting only paired nucleotides \cite{Aaron}. The figure indicates that viral RNA molecules tend to have shorter maximum ladder distances than random sequences. The secondary structure of a large ssRNA molecule is not fixed. For a vRNA molecule, thousands of alternative secondary structures may be found with free energies that differ by less than $k_BT$ from the minimum free energy state. The structures shown in Fig.~\ref{fig:Sec} thus should only be viewed as representative of a large family of secondary structures with nearly the same free energy. The entropy associated with this quasi-degeneracy plays an important role in the theory.

\section{Second Virial Coefficient $V_1$}

\begin{figure}[htb]
\centering
\includegraphics[width=0.2\textwidth]{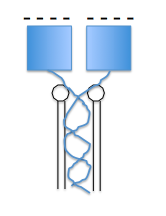}\\
\caption{\label{VTT} Attraction between capsid proteins can be mediated by the association of positively charged tail charges with negatively charged RNA segments.}\end{figure}

The second virial coefficient $V_1$ of saturated aggregates plays the role of an effective temperature in the model. Here, we estimate different contributions to  $V_1$. We estimate the electrostatic contribution $V_e$ to $V_1$ as the second virial coefficient of a saline solution containing spherical particles of diameter $D$ and charge $Q$. For particle radii larger than the Debye screening length, so for $\kappa D \gg 1$, $V_e=\frac{Q^2l_B}{2\kappa^3 D }$ in the DH approximation in a monovalent salt solution \cite{Philipse}. For CCMV the head-group charge $Qe$ decreases from about $12e$ to about $10e$ as the pH is decreases from $pH\simeq7$, when empty capsids do not form, towards the isoelectric point around $pH\simeq5$, when empty capsids do readily assemble \cite{Vega}. For $D$ equal to \unit[3]{nm}, $\frac{Q^2l_B}{2\kappa^3 D }$ decreases from about $\unit[30]{nm^3}$ to about $\unit[20]{nm^3}$. The next term, $-V_a$, is due to the attractive interactions between CPs. The main contribution is hydrophobic attraction, which does not depend on pH. Since in the absence of RNA, empty capsids do not assemble at neutral pH but do assemble as the pH is reduced, the magnitude $V_a$ of the attractive interaction is estimated to be between $\unit[20]-\unit[30]{nm^3}$. Finally, when RNA is present, the RNA molecule can mediate attractive interactions by bridging (see Fig.\ref{VTT}) with a contribution $-V_T$to the second virial coefficient. For CCMV, the tail charge is not significantly dependent on the pH, in which case $V_T$ is not expected to depend on the pH, though it should depend on the ionic strength.
\begin{equation}\label{fmix2}
 V_1\simeq V_e-V_a -V_T
\end{equation}

The sum of the first two terms is positive for neutral pH. In order for $V_1$ to be negative under reduced pH, when CCMV virions form, but not for neutral pH when stable CCMV virions do not form, $V_T$ should be in the range of a few times $\unit{nm^3}$. Given this complexity, instead of attempting to calculate $V_1$ it may be more practical to directly measure $V_1$ through the equation of state of a mixture of CPs and nanometer length ssRNA segments with a concentration ratio $X=1$.

\section{Aqueous Electrostatics}

In this section we discuss Poisson-Boltzmann (PB) theory as applied to the proposed model  (see, e.g., \cite[Chapter 3]{ColloidalDomain}, \cite[Chapter 10]{Doi_SoftMatter}.
\subsection{Charged Spheres}
As a simple model for the interior of the virion, assume a uniformly charged sphere of charge $Qe$ and radius $R$ in a solution of monovalent salt ions with concentration $c_s$. The sphere is permeable to the salt ions. The charging parameter is defined as $\alpha(R)=\frac{|Q|}{V(R)c_s}$ with $V(R)=(4/3)\pi R^3$. The PB electrostatic free energy of the charged sphere equals:
\begin{equation}\label{clp1b}
\begin{split}
\frac{F_{\mathrm{PB}}(\alpha)}{k_BT V(R)c_s} = & \ 2\left( 1-{\sqrt{1+\alpha^2}}\medspace \right)+\\& +\alpha\ln \left( \frac{ \alpha + {\sqrt{1 + \alpha^2}}}{-\alpha+{\sqrt{1+\alpha^2}}}\right)
\end{split}
\end{equation}
The electrical potentials inside and outside the sphere are nearly constant but at the surface of the sphere there is a potential dros by an amount $\Delta \Phi_D$ given by:
\begin{equation}\label{clq1b}
\frac{ e }{k_BT}\Delta\Phi_D=-\ln \left(\alpha+{\sqrt{1+\alpha^2}}\medspace \right) \ ,
\end{equation}
which is a version of the Donnan Potential. The electrostatic free energy in the weak-charging regime $\alpha(R)\ll1$ is quadratic in $\alpha(R)$:

\begin{equation}\label{clp4b}
\frac{ F_{\mathrm{PB}}(k_BT\alpha)}{Vc_s} \sim \alpha^2 \quad\quad  \alpha(R)\ll1
\end{equation}
In the strong charging regime $\alpha(R)\gg1$, it adopts the form
\begin{equation}\label{clp4b}
\frac{F_{\mathrm{PB}}(k_BT \alpha)}{Vc_s} \sim 2\alpha \ln \alpha \quad\quad \alpha(R) \gg 1
\end{equation}
This expression can be viewed as the entropic free energy of the counterions of the macromolecule. In the strong charging regime, the Donnan potential is of the order of $\frac{k_BT}{e}$. The electrostatic free energy difference of a monovalent counterion inside and outside the sphere is thus comparable to the thermal energy in the strong charging regime.

\subsection{Charged Shells}
The PB electrostatic free energy per unit area of a charged shell of radius $R$ with surface charge density $e\sigma$ equals (see, e.g., \cite[Chapter 3]{ColloidalDomain})
\begin{equation}\label{919d} \frac{ F_{\mathrm{PB}}}{k_B T A} = 2 \sigma \left[\lambda \kappa - \sqrt{1 + (\lambda
\kappa)^2} + \ln \frac{1 + \sqrt{1+(\lambda \kappa)^2}}{\lambda
\kappa} \right] \ .  \end{equation}
in the large R limit. Here, $\lambda= 1/(\pi \sigma l_B)$ is the \textit{Gouy-Chapman (GC) length} and $\kappa$ is again the inverse of the Debye screening length. In the weak-charging regime $\lambda \kappa  \gg  1$,  formula (\ref{919d}) simplifies to
\begin{equation}
\frac{F_{\mathrm{PB}}}{k_B T A} \simeq \frac{\sigma^2l_B}{\kappa}
\end{equation}
while in the strong charging limit, $\lambda \kappa \ll 1$,
\begin{equation}\label{919}
\frac{F_{\mathrm{PB}}}{k_B T A} \simeq 2 \sigma \left[ \ln\frac{2}{\lambda\kappa}-1 \right] \ .
\end{equation}
For smaller $R$, curvature effects do contribute to the surface electrostatic free energy  \cite{Lekkerkerker_1989} but this can be neglected in our case.

The coefficient ``two" in front of Eq. (\ref{919}) may seem strange on first glance: $Q_h\rho$ is the 2D concentration of counterions, so one could have expected just an ideal gas of counterions contributing a surface osmotic pressure $\Pi=k_BTQ_h\rho$ (or lateral pressure). In fact, counterions are localized around the surface within layer whose thickness is the GC length, $\lambda = 1/\pi l_B\sigma$, which depends on the surface area through the surface charge density $\sigma$. The ideal gas of counterions is confined to a volume $\lambda$A $\propto A^2$; it is this $A^2$ that produces the factor of two upon differentiation with respect to area when the osmotic pressure is computed. Physically, it means that the lateral pressure of counterions is enhanced by the fact that the counterions are attracted to the surface.

\subsection{Weak and strong charging regimes of confined polyelectrolyte molecules.}

\begin{figure}[htb]
\centering
\includegraphics[width=0.45\textwidth]{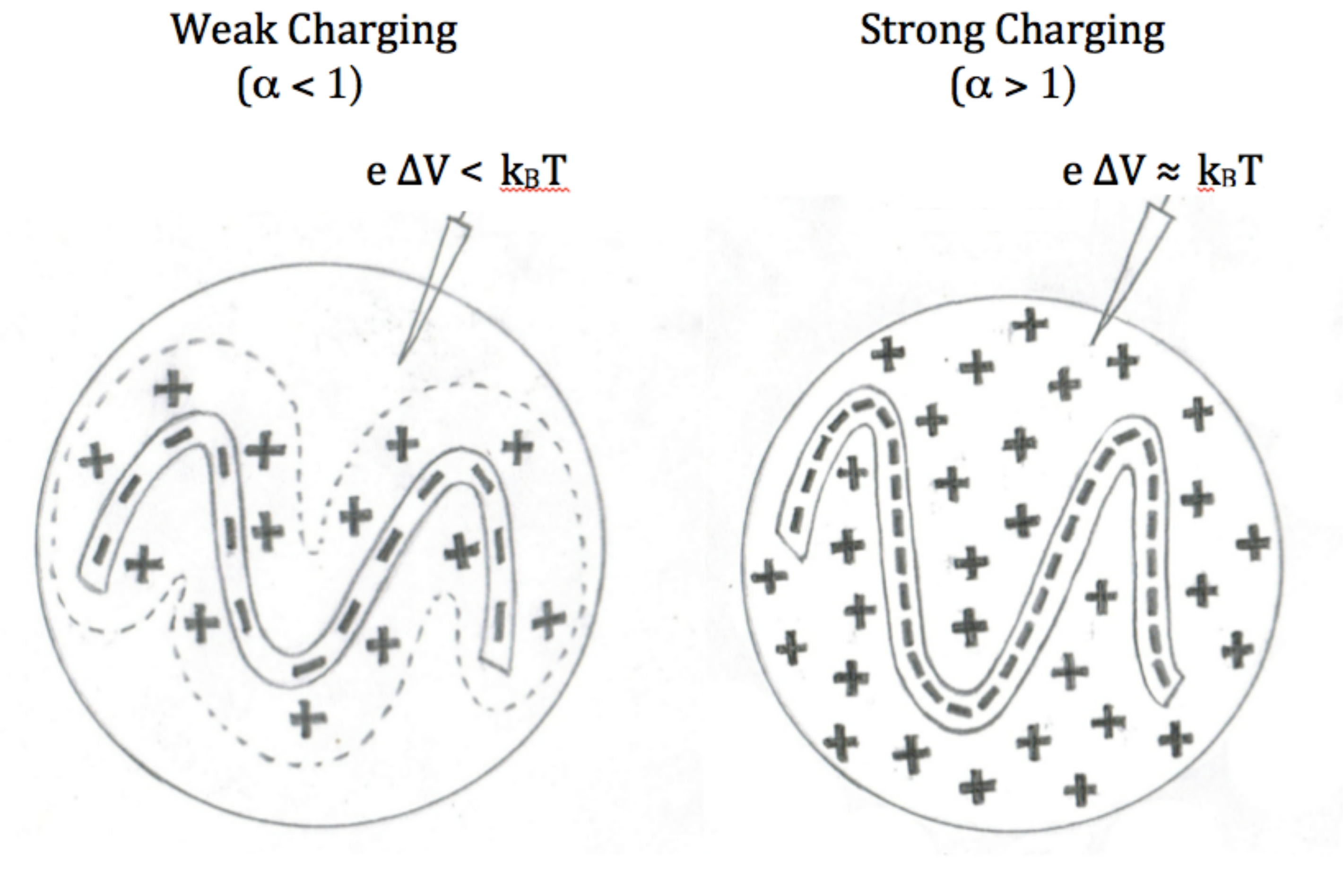}\\
\caption{\label{Donnan} Weak and strong charging regimes.}\end{figure}

The uniform charge model is not quite adequate in the weak-charging regime to describe a confined polyelectrolyte molecule. Let $l$ be the persistence length of a highly charged polyelectrolyte molecules with total length $L=Nl$ and total charge $Qe$. Assume that molecule is confined inside a sphere of radius R.  The charging parameter is then $\alpha(R)=\frac{|Q|}{V(R)c_s}$. In the strong charging regime $\alpha  \gg  1$ the counterions are distributed relatively uniformly over the globule (see Fig.~\ref{Donnan}).

The electrostatic free energy of the molecule in the strong charging regime $\alpha(R)\gg1$ can be obtained by applying Eq. (\ref{clp4b}):
\begin{equation}\label{clp4b}
\frac{ F_{\mathrm{PB}}}{k_BT} \sim 2Q \ln \left( \frac{Q}{V(R)c_s} \right)\quad\quad\alpha(R) \gg 1
\end{equation}
The electrostatic free energy of the polyelectrolyte molecule in the weak-charging regime $\alpha(R) \ll 1$ can not be obtained from the charged sphere expression because the counterions of the polyelectrolyte molecule are no longer uniformly distributed but confined to a tube surrounding the polyelectrolyte molecule with a radius that is of the order of the screening length (see Fig.~\ref{Donnan}). The DH expression for the electrostatic free energy of a polyelectrolyte molecule was obtained by Onsager \cite{Onsager_LC} and it has the form

\begin{equation}\label{clp4b}
\frac{ F_{\mathrm{PB}}}{k_BT} \sim \left( \frac{l^2}{\kappa} \right) \frac{N^2}{R^3} \quad\quad \alpha(R) \ll 1
\end{equation}
The actual validity condition of the Onsager/DH result is that the volume $L\kappa^{-2}$ of the Debye screening cloud surrounding the $N$ segments has to be less than the volume of the sphere, but this reduces to $\alpha(R) \ll 1$.

\section{Assembly of Virions: Law of Mass Action}
 In this section we appy the Law of Mass Action (LMA) to virion assembly. Recall that the LMA is known to be a good description for the assembly of \textit{empty} capsids. The starting point is the free energy of a dilute solution of proteins and RNA molecules expressed as:
\begin{equation}\label{MA}
\begin{split}
\frac{F}{k_BT} & =\phi_f\ln \left(\frac{\phi_f}{c_0} \right)+\\ +&\sum_{p=1}^{N}\left( [C(p)]\ln \left( \frac{[C(p)]}{c_0} \right) -E(p)[C(p)]\right) \ .
\end{split}
\end{equation}
The first term is the solution entropic free energy of free CPs, with $\phi_f$ the concentration of free CPs and $c_0\sim 1/D^3$ the dense-packing concentration of virions. The first part of the second term is the solution entropic free energy of aggregates composed of one vRNA molecule with a varying number of proteins. $[C(p)]$ is the concentration of vRNA molecules associated with $p$ CP molecules. $N$ is the maximum number of CPs per aggregate. In last term, $E(p)$ is the association energy of a $p$-protein aggregate. This free energy is to be minimized subject to the conservation laws for the RNA and the CP molecules:
\begin{equation}\label{C}
\begin{split}
&\phi_{\mathrm{RNA}}=[\mathrm{RNA}]_f+\sum_{p=1}^{N}[C(p)]\\
&\phi_{\mathrm{CP}}=\phi_f+\sum_{p=1}^{N} p[C(p)]
\end{split}
\end{equation}
Here, $\phi_{\mathrm{RNA}}$ is the total RNA concentration, $[\mathrm{RNA}]_f$ the concentration of free RNA molecules not associated with CPs while $\phi_{\mathrm{CP}}$ is the total concentration of CP molecules. Minimization with respect to $C(p)$ subject to these constraints leads to
\begin{equation}\label{D}
[C(p)]=[\mathrm{RNA}]_f\left(\frac{\phi_f}{c_0}\right)^p \exp \left(E(p)/k_BT \right) \ ,
\end{equation}
which can be viewed as expressing the LMA or as expressing the Boltzmann distribution in the Grand Canonical Ensemble. The concentration $\phi_f$ of free CPs must obey the condition
\begin{equation}\label{E2}
\phi_{\mathrm{CP}}=\phi_f+\phi_{\mathrm{RNA}}\left< p \right>
\end{equation}
Here, $\left< p \right>$ is the expectation value of the number of CPs per RNA molecule obtained from the probability distribution $[C(p)]/\phi_{\mathrm{CP}}$.

First consider the simplest case: a solution that contains only free CPs, free vRNA molecules and assembled virions. The self-consistency condition Eq. (\ref{E2}) then reads:
\begin{equation}\label{E}
\bar{\phi}_{\mathrm{CP}}= \bar{\phi}_f+M\bar{\phi}_{\mathrm{RNA}}\frac{\bar{\phi}_f^M \exp \left( E_v/k_BT \right)}{1+\bar{\phi}_f^M\exp \left(E_v/k_BT \right))} \ ,
\end{equation}
where $E_v$ is the association energy of a capsid and where all concentrations with a bar have been made dimensionless by dividing by $c_0$.

The solution of this equation is shown in Fig.12 of the main text. For $\bar{\phi}_f$ less than $\phi^{\ast}=\exp \left(-(E_v/M)/k_BT \right)$ the second term on right-hand side of the equation can be neglected so $\bar{\phi}_f\simeq \bar{\phi}_{\mathrm{CP}}$: the CPs are nearly all free in solution, as are the RNA molecules (the uniform phase of Fig.3).  When $\bar{\phi}_{\mathrm{CP}}$ approaches $\phi^{\ast}=\exp \left( -(E_v/M)/k_BT \right)$ the second term starts to rise very sharply. For larger $\bar{\phi}_{\mathrm{CP}}$, $\phi_f$ remains pegged at $\phi^{\ast}$ while the virion concentration grows linearly as $\bar{C}(M)\simeq(\bar{\phi}_{\mathrm{CP}}-\phi^{\ast})/M$. This stops when $\bar{\phi}_{\mathrm{CP}}$ exceeds $\bar{\phi}^{\ast}+M\bar{\phi}_{\mathrm{RNA}}$ where nearly all RNA molecules have been encapsidated. The concentration of free CPs then starts to grow again as $\bar{\phi}_{\mathrm{CP}}$ is increased beyond this point. The optimal concentration ratio between the vRNA and CP concentration is thus the stoichiometric ratio of the virion.

The CP concentration $\bar{\phi}_{\mathrm{CP}}$ is the essential thermodynamic variable that regulates assembly in the LMA description. The boundary for virion assembly as a function of thermodynamic parameters is determined by the condition $\bar{\phi}_{\mathrm{CP}}\simeq\phi^{\ast}$ or
\begin{equation}\label{E}
E_v=-Mk_BT\ln\bar{\phi}_{\mathrm{CP}}
\end{equation}
which corresponds to equating the CP chemical potential in solution with that of a CP that is part of a virion.

To arrive at an explicit expression for the transition line in the $\varepsilon-u$ plane we need an expression for the association energy $E_v$ of an assembled virion composed of one vRNA molecule and $M$ CPs. In a naive model, $E_v(\varepsilon,u)$ could be written as the sum of three parts:
\begin{equation}\label{F}
E_v(\varepsilon,u)=\left(\varepsilon+\frac{1}{2}z_v u\right)M-\Delta F_v
\end{equation}
In the first term, $\varepsilon$ is again the association free energy between a CP tail and a vRNA molecule while $z_v$ is the mean number of neighbors of a CP incorporated in the ordered capsid of a virion (between 5 and 6), while $u$ is the CP-CP pairing energy. The last term, $\Delta F_v$, is the change in free energy of the vRNA molecule before and after encapsidation, not counting the CP/RNA association energy. It includes the free energy associated with RNA condensation - excluding the RNA/CP binding energy - and it may be negative or positive but it is not expected to depend on $\varepsilon$ and $u$. Under this ansatz, the transition line for the onset of virion assembly is a straight line in the $\varepsilon-u$ plane \footnote{The condition that there is no empty capsid assembly is $M\varepsilon -\Delta F_v+k_BT\ln\bar{\phi}_{\mathrm{RNA}}>0$}.

One can include in the LMA description assembly intermediates in the form of partially assembled shells. The energy cost of the edge of a partial shell is of the order of $u$ times the number of CPs that constitute the edge of a partial shell. The maximum value of the edge length is of the order of $M^{1/2}$. It follows from the fact that $u M^{1/2}$ is large compared to $k_BT$ when $u$ is of the order of a few $k_BT$ that, under conditions of thermodynamic equilibrium, the concentration of assembly intermediates is negligible as compared to that of free CPs and assembled virions. This argument is the same as the reason why the concentration of partial shells is negligible during the assembly of empty capsids \cite{Zandi}.

\section{Chemical Potential and Disproportionation}

If $\varepsilon/k_BT$ is not large compared to one then the mean segment occupancy probability $\left< x \right>$ must be less than $X$ since some CPs now will remain free in solution, say at a concentration of $\phi_f$. The condition of phase equilibrium between these free CPs and those that are part of an aggregate is that they must have the same CP chemical potential $\mu$. The concentration $\phi_f$ of free CPs is given by
\begin{equation}
\begin{split}
\phi_f=\phi_{\mathrm{CP}}\left(1-\frac{ \left< x \right> }{X}\right)
\end{split}
\end{equation}
so $\mu=k_BT\ln(\phi_{\mathrm{CP}}\left(1-\frac{ \left< x \right> }{X}\right)/c_0)$.

Equating $\mu$ to the chemical potential inside an aggregate gives:
\begin{equation}\label{CT}
\begin{split}
k_BT\ln\left(\phi_{\mathrm{CP}}\left(1-\frac{ \left< x \right> }{X}\right)/c_0\right)+\varepsilon=\frac{1}{N}\frac{\partial F(x)}{\partial x}
\end{split}
\end{equation}
Here, $F(x)$ is the value of the Flory variational free energy $F(R,S,x)$, see Eq.II.1, after minimization with respect to $R$ and $S$ for fixed $x$. The argument of $F(x)$ - the occupancy of a particular aggregate - must be distinguished here from $ \left< x \right> $, its average over all aggregates. Though $ \left< x \right> $ necessarily is less than $X$, individual aggregates can have an occupancy that exceeds $X$.

\begin{figure}[htb1]
\centering
\includegraphics[width=0.45\textwidth]{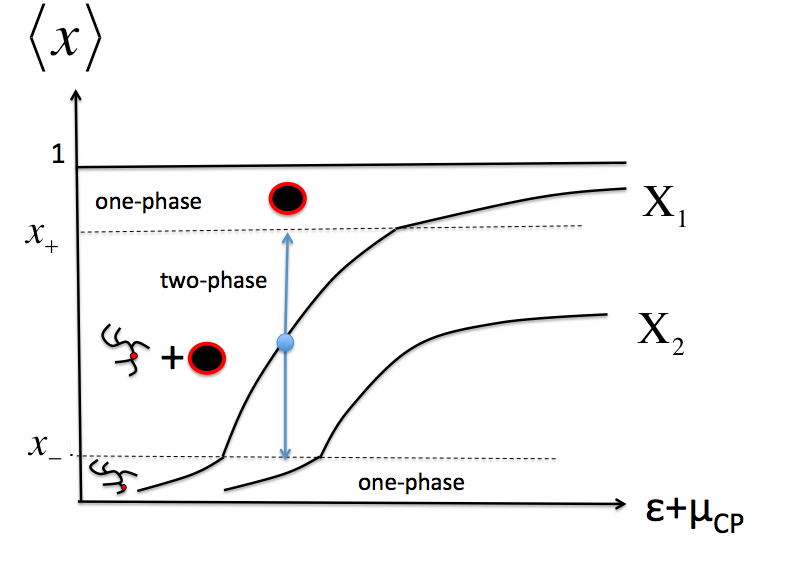}\\
\caption{\label{fig:Segregation15} Disproportionation of a solution of RNA and capsid proteins. The vertical axis $\left< x \right>$ is the probability that an RNA segment is associated with a capsid protein (CP). The horizontal axis is the chemical potential $\mu_{CP}$ of the CPs in the absence of RNA pluse the tail-RNA binding energy. The two solid lines are curves with the same fixed second virial coefficient $V_1<V_c$ but different macroscopic mixing ratios $X$, with $X=1$ a saturated aggregate. If $ \left< x \right>$ is in the interval between $x_1$ and $x_2$ there is disproportionation into CP poor and CP rich aggregates, as shown by the vertical blue line. Note that the coexistence line is smeared. If $X$ is reduced from $X_1$ to $X_2<x_2$, then the solution remains disproportionated for large values of $\epsilon+\mu_{CP}$. }\end{figure}

It is convenient to replace $\mu$ with the CP chemical potential $\mu_{\mathrm{CP}}=k_BT\ln(\phi_{\mathrm{CP}}/c_0)$ of CPs in the absence of vRNA molecules. The latter quantity, but not the former, is under experimental control (it is roughly in the range of $-6$ for the for the \textit{in-vitro} experiments discussed in the conclusion. In terms of $\mu_{\mathrm{CP}}$, the condition of phase equilibrium for finite $\varepsilon$ translates to
\begin{equation}\label{FE}
\begin{split}
\frac{ \left< x \right> }{X}=1-\exp\left(-\frac{1}{k_BT}\left[\mu_{\mathrm{CP}}+\varepsilon-\frac{1}{N}\frac{\partial F(x)}{\partial x}\right]\right)
\end{split}
\end{equation}
We now can use Eq.\ref{FE} to follow the disproportionation process as a function of the mixing ratio $X$ and $\epsilon+\mu_{\mathrm{CP}}$ using the common tangent method. The result shown in Fig. \ref{fig:Segregation15}.
The two solid lines are loci with the same the second virial coefficient $V_1$ fixed at a negative value below the critical point with a corresponding disproportionation interval is $[x_1,x_2]$. For the top curve, the macroscopic mixing ratio $X_1$ is in the interval $[x_2,1]$. For $\mu_{\mathrm{CP}}+\varepsilon \gg k_BT$, $ \left< x \right>$ approaches $X_1$.If $\mu_{\mathrm{CP}}+\varepsilon$ is reduced, then so does the mean occupancy $ \left< x \right> $. Disproportionation starts when $ \left< x \right> $ drops below $x_2$. The system disproportionates between $x_1$ and $x_2$, according to the usual tie rule (see Fig. \ref{fig:Segregation15}). Note the kink in the dependence of $ \left< x \right> $ on $\mu_{\mathrm{CP}}+\varepsilon$ at $ \left< x \right> =x_2$ and $ \left< x \right> =x_s$. For small $\mu_{\mathrm{CP}}+\varepsilon$, when all globules have converted to coils, the mean occupancy eventually goes to zero. If the mixing ratio $X$ is reduced to a value $X_2<x_2$ then the system remains  in the two-phase region for large $(\mu_{\mathrm{CP}}+\varepsilon)$. The physical reason for the ``smearing'' of the transition, as compared to conventional phase separation, is that the experimentally accessible chemical potential $(\mu_{\mathrm{CP}}+\varepsilon)$ is not the actual chemical potential.

\section{Surface Tension}
An aspect of the globule state that is not covered by Flory theory  concerns the \emph{surface tension} of the vRNA globule, denoted by $\gamma_0$. The globule surface tension is related to the so-called ``blob size'' - the correlation length - by the scaling relation $\gamma_0 \approx{k_BT/{\xi_g}^2}$ (see refs. \cite[Section 20]{RedBook} and \cite[Section 3.3.2]{RubinsteinColby}). We estimate $\gamma_0$ as the characteristic condensation energy per segment $k_BT{V^2}/W$ divided by the area per segment $l^2$. For the previously estimated values of $V$, $W$, and $l$ used, $\gamma_0$ would be in the range of $\unit[]{erg/cm^2}$ (about an order of magnitude less than that of a typical liquid). Because of this surface tension, polymer globules tend to aggregate when they come in contact. A fused aggregate globule composed of two polymers has a lower surface area than two separate globules. A coil-to-globule transition is likely to induce macroscopic phase separation of a solution of vRNA molecules when vRNA globule would fuse under the action of surface tension. Under physiological conditions, vRNA molecules do not aggregate in this fashion. The natural interpretation is that the second virial co- efficient V is positive for vRNA molecules under physiological conditions. Saturated aggregates and assembled virions should not aggregate either. This is indeed prevented by the fact that the total charge of CPs is small so CP-RNA aggregates remain charged. Electrostatic repulsion can then prevent aggregation.

\section{Optimal Mixing Ratio}
\label{VA}

In this section we compute the optimal mixing ratio, or OMR, using the Law of Mass Action (LMA) for solutions in thermal equilibrium that contain free CPs, free RNA molecules, virions, and saturated aggregates. Minimization of the solution free energy of Section IV gives
\begin{equation}\label{4991}
\begin{split}
&[\phi_{\mathrm{CP}}]=[\phi_{\mathrm{CP}}]_f+\\&+[\phi_{\mathrm{RNA}}]\left(\frac{M{[\phi_{\mathrm{CP}}]_f}^M e^{E_v/k_BT} +N{[\phi_{\mathrm{CP}}]_f}^N e^{E_{pv}/k_BT}}{1+[{\phi_{\mathrm{CP}}]_f}^M e^{E_v/k_BT} +{[\phi_{\mathrm{CP}}]_f}^N e^{E_{pv}/k_BT}} \right)
\end{split}
\end{equation}
All concentrations are here dimensionless. The quantity inside the large brackets is the expectation value of the number of CPs associated with an RNA molecule in the grand-canonical ensemble, with $k_BT\ln[\phi]_f$ the chemical potential of the CPs. $E_v$ and $E_{pv}$ are the assembly free energies of an M CP virion, respectively, an N CP provirion. Consider that the macroscopic CP to RNA mixing ratio $X=N[\mathrm{CP}]/[\mathrm{RNA}]$ is less than one. The stoichiometric mixing ratio that corresponds to virion assembly is $X^{\ast}=M/N$. The OMR is the value of $X$ that maximizes the fraction $E(X)$ of RNA molecules that is part of a virion where

\begin{equation}\label{4992}
\begin{split}
&E(X)=\frac{{(z_v[\phi_{\mathrm{CP}}]_f})^M}{\left(1+(z_v[{\phi_{\mathrm{CP}}]_f)}^M+{(z_{pv}[\phi_{\mathrm{CP}}]_f)}^N\right)}
\end{split}
\end{equation}
where notations $z_v=\exp \left( E_v/k_BT M \right)$ is the Boltzmann factor per CP in the virion state and $z_{pv}=\exp\left( E_{pv}/Nk_BT \right)$ the CP Boltzmann factor in the provirion state. The transition from a provirion to a virion state as a function of the CP-CP binding energy $u$ requires that the CP/RNA binding energy $\varepsilon$ is large compared to $k_BT$, which means that both $z_v$ and $z_{pv}$ are large compared to one. The conservation law for CPs then reduces to

\begin{equation}\label{4993}
\begin{split}
&X\simeq \frac{(M/N){(z_{pv}[\phi_{\mathrm{CP}}]_f)}^M+{(z_v[\phi_{\mathrm{CP}}]_f)}^N}{1+{z_v[\phi_{\mathrm{CP}}]_f}^M+{(z_{pv}[\phi_{\mathrm{CP}}]_f)}^N}
\end{split}
\end{equation}
which simplifies to

\begin{equation}\label{7894}
\begin{split}
&X\simeq {(X^{\ast}-X)({[\phi_{\mathrm{CP}}]_f z_V})^M+(1-X)({[\phi_{\mathrm{CP}}]_f z_{pv}})^N}
\end{split}
\end{equation}
where we introduced the notation $X^{\ast}=M/N$ for the stoichiometric mixing ratio.

For mixing ratios $X$ below the stoichiometric ratio $X^{\ast}$, both terms on the right hand side of Eq. (\ref{7894}) are positive. If both ${[\phi_{\mathrm{CP}}]_f z_v}$ and ${[\phi_{\mathrm{CP}}]_f z_{pv}}$ are less than one then the equation has no solution since the left hand side is of the order of one while both terms on the right hand side are small compared to one (assuming that both $M$ and $N$ are large compared to one). If $z_{pv}$ exceeds $z_{p}$ then, with increasing $[\phi_{\mathrm{CP}}]$, the factor ${[\phi_{\mathrm{CP}}]_f z_{pv}}$ will reach one first at which point the first term on the right hand side is of the order of one. Since ${[\phi_{\mathrm{CP}}]_f z_v}$ will be less than one, the second term is small compared to one when raised to the power $M$. The equation can be solved and the concentration of free CPs is
\begin{equation}\label{4995}
\begin{split}
z_{pv}{{[\phi_{\mathrm{CP}}]_f}}\simeq &\left(\frac{X}{1-X}\right)^{1/N}
\end{split}
\end{equation}
Note that the right hand side is close to one unless $X$ is close to zero or one. There are practically no virions in this case. If $z_{pv}$ is less than $z_{p}$, then the roles of virions and provirions are exchanged. The solution of Eq. (\ref{7894}) now leads to

\begin{equation}\label{4996}
\begin{split}
z_{v}{{[\phi_{\mathrm{CP}}]_f}}\simeq &\left(\frac{X}{X^{\ast}-X}\right)^{1/M}
\end{split}
\end{equation}
Inserting this in to the expression for the fraction of encapsidated RNA molecules gives

\begin{equation}\label{4997}
\begin{split}
E(X)\simeq&X/X^{\ast}\quad\quad X<X^{\ast}
\end{split}
\end{equation}
Now, there are practically no provirions. The provirion to virion transition line is thus given by $z_{pv}=z_{p}$ or by $E_v/M = E_{pv}/N$. This means that, at the transition point, the assembly energy \textit{per CP} must be the same for virions and provirions. One way to understand this result is by noting that it is equivalent to demanding that the chemical potential of a CP that is part of a virion must be the same as one that is part of a provirion state at the transition point.

Next consider the case that the mixing ratio exceeds the stoichiometric ratio (so $X>X^{\ast}$). The excess CPs compete with the CPs that are part of a virion for excess to RNA. Now the first term of Eq. (\ref{7894}) is negative. If $z_{pv}$ exceeds $z_{v}$ then with increasing $[\phi_{\mathrm{CP}}$ the factor $({[\phi_{\mathrm{CP}}]_f z_{pv}})$ again reaches one first. The positive second provirion term exceeds the negative virion term and the same result ensues as before. However, in the opposite case that $z_{v}$ exceeds $z_{pv}$ no solution appears when $({[\phi_{\mathrm{CP}}]_f z_{pv}})$ reaches one because the virion term is negative. Instead, $[\phi_{\mathrm{CP}}$ must be increased further until the provirion term cancels the virion. This happens when

\begin{equation}\label{789}
\begin{split}
 {(X-X^{\ast})({[\phi_{\mathrm{CP}}]_f z_V})^M\simeq(1-X)({[\phi_{\mathrm{CP}}]_f z_{pv}})^N}
\end{split}
\end{equation}

The solution contains a mixture of virions and provirions and only few free RNAs or CPs. From this condition, the fraction of encapsidated RNA molecules follows directly

\begin{equation}\label{499}
\begin{split}
&E(X)=\frac{(1-X)}{(1-X^{\ast})}\quad\quad X>X^{\ast}
\end{split}
\end{equation}
The function $E(X)$ has a cusp maximum at $X^{\ast}=M/N$, the stoichiometric ratio for the assembly of a virion. We conclude that according to the LMA, $X^{\ast}$ is the optimal mixing ratio for virion assembly under conditions of full thermodynamic equilibrium and binding energies large compared to the thermal energy.

\section{Surface Segregation}
\label{SC}

\begin{figure}[htb]
\centering
\includegraphics[width=0.45\textwidth]{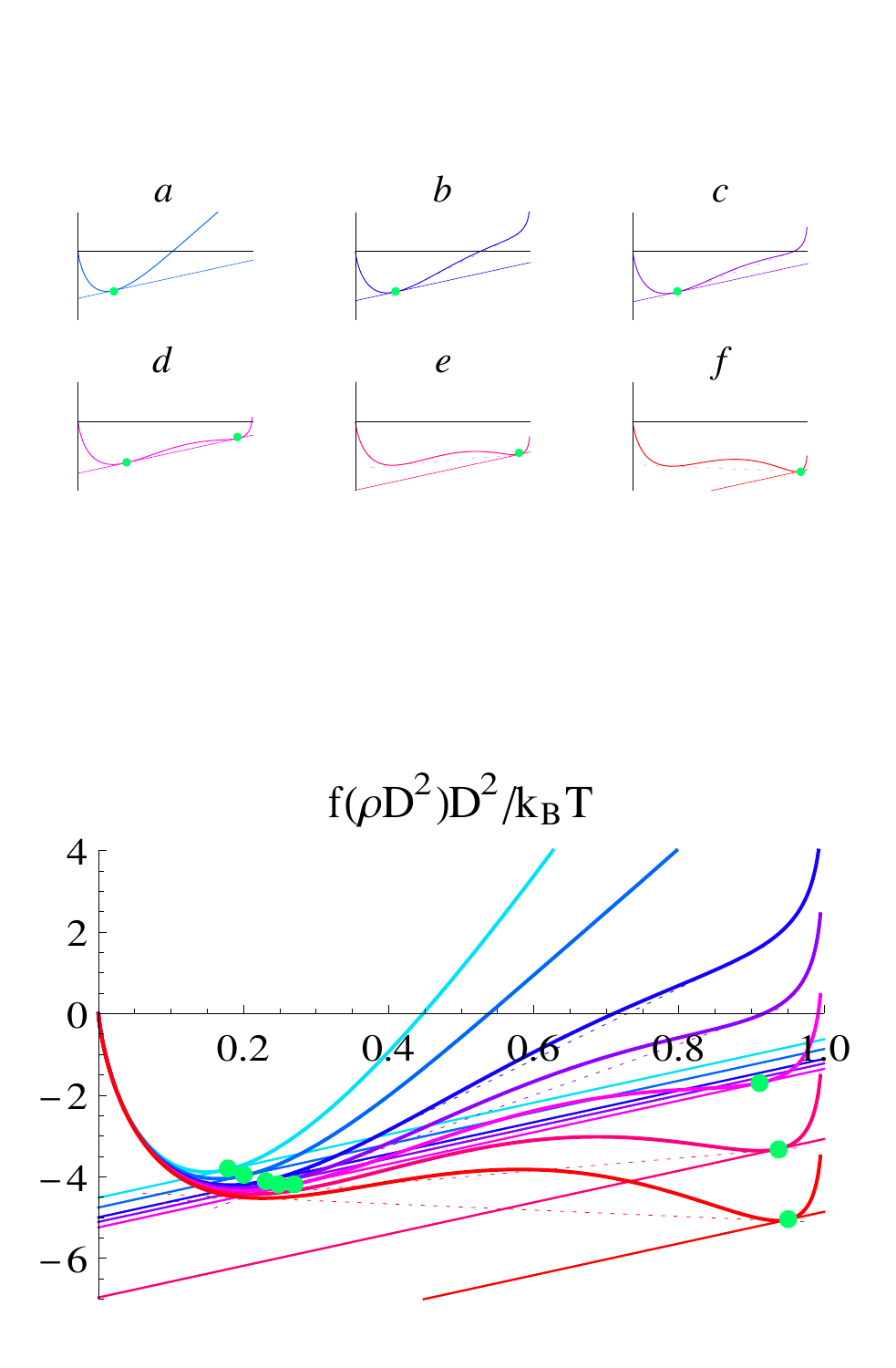}\\
\caption{\label{fig:Surface} Bottom panel: surface free energy as a function of CP surface density $\rho$ for increasingly negative values of $B$.}\end{figure}

In this section we discuss surface segregation. Assume that CPs on the surface of the globule are described by the modified van der Waals system of the proposed model:
\begin{equation}\label{456}
\begin{split}
{{f}(\rho_2,\psi)}/{k_BT}  = &\rho_2\ln\left(\frac{\rho_2 D^2}{1-{\rho_2 D^2}}\right)+2\rho_2 Q \ln\left(\frac{\rho_2 Ql_B}{\kappa}\right)+\\&+B\rho_2^2
\ . \end{split} \end{equation}
The surface energy density is plotted in Fig.~\ref{fig:Surface} as a function of the area density for increasingly negative values of $B$. %
The surface CPs are in chemical equilibrium with the CPs in the interior of the globule, which have a chemical potential $\mu\simeq\gamma_0D^2$. For $|B|$ small compared to $B^{\ast} \approx -k_B T D^2 Q$, the repulsive electrostatic interactions dominate. The condition for phase equilibrium with the bulk produces a relatively low surface density (indicated by the small solid dots in panels a - d). The solid dots in the main figure, which all indicate states with low surface density, form an almost continuous string.  As $B$ becomes more negative, attractive interactions generate a negative curvature for intermediate surface densities. There is a specific value $B^{\ast}$, of the order of a few times $Q D^2$,  where the interior coexists with both the low density surface density state and the high surface density state. The solid tangent line coincides with the dashed common tangent.  For larger values of $|B|$, only this state remains.

\begin{table}
\caption{Table of Latin symbols}
\begin{tabular}{||c|c||}

\hhline{|t:=:t:=:t|}
$A$ & surface area \\
$B_{\psi}$ & second virial coefficient \\
$B^{\ast}$ & value second virial coefficient \\ & at critical point. \\
$b$ & spacing between charges \\
$c_0$ & 1/virion volume\\
$c_s$ & salt concentration \\
$C{p}$ & solution concentration of\\& aggregates containing p proteins\\
$D^2$ & excluded CP area \\ &  on capsid surface \\
$E_v$ & cohesion energy of a virion\\
$F$ & solution free energy\\
$F_F(R)$ & Flory free energy of a vRNA molecule\\
$F_{PB}$ & PB Electrostatic free energy\\
$F(R,x)$ & Variational free energy of aggregate\\
$F_2$ & surface free energy\\
$f(\rho_2)$ & surface free energy area density\\
$1/g$ & average \# segments between \\ & branch points \\
$K_H$ & Helfrich bending energy of a CP shell \\
$l$ & length of one segment \\
$l_B = \frac{e^2}{\varepsilon k_B T}$ & Bjerrum length \\
$M$ &\# CPs per T=3 virion (180) \\
$N$ & maximum \# CPs per RNA molecule \\
$n$ & number of CPs whose tail groups are \\ &in contact with the vRNA molecule\\
$Q^{\ast}$ & Effective macroion charge. \\
$Q_t>0$ & charge of a CP tail group in \\ & units of elementary charge $e$ \\
$-Q_h<0$ & charge of a CP head group in \\ & units of elementary charge $e$ \\
$Q=Q_t=Q_h$ & approximated value \\ &of the tail and head group charges\\
$[{RNA}]_f$ & solution concentration of\\& CP-free vRNA molecules\\
$R$ & radius of gyration\\& of a vRNA molecule\\
$R_s$ & mean radius of curvature\\& of the shell\\
$R_c$ & radius of a T=3 capsid\\
$R_g$ & equilibrium radius of gyration\\& of a vRNA molecule\\
$S$ & spanning distance of branched \\ & RNA in units of segments \\
$u$ & affinity between CP head groups\\
$V$ & second virial c-t of RNA segments\\
$V_a$ & contribution to second virial c-t of \\ &aggregates due to CP-CP attraction\\
$V_e$ & contribution to second virial c-t of \\ &aggregates due to CP-CP \\ &electrostatic repulsion\\
$V_{\mathrm{eff}}(x)$ & second virial c-t of \\ & an aggregate\\
$V_{T}$ & second virial c-t \\ & of RNA segments due to \\ & CP tail-induced attraction\\
$V_{1}$ & second virial c-t  of a saturated aggregate\\
$W$ & third virial c-t of RNA segments\\
$W_{\mathrm{eff}}(x)$ & third virial c-t of \\ & an aggregate\\
$X$ & concentration ratio of CPs and\\& vRNA segments (macroscopic)\\
$x$ & ratio of the number of CPs and\\&  vRNA segments of an \\&aggregate (microscopic)\\

\hhline{|b:=:b:=:b|}
\end{tabular}
\end{table}

\begin{table}
\caption{Table of Greek symbols}
\begin{tabular}{||c|c||}

\hhline{|t:=:t:=:t|}
$\alpha = \frac{Q^{\ast}}{2 c_s V}$ & charging parameter \\
$\alpha \gg 1$  & strongly charged regime \\
$\alpha \ll 1$ & weakly charged regime \\\
$\gamma_0$ & Surface tension of a\\& vRNA globule in poor solvent\\
$\Delta F_v$ & change in free energy of the vRNA\\& molecule upon assembly\\
$\Delta\psi$ & angular width of the \\& directional CP-CP pair potential.\\
$\varepsilon$ & affinity between RNA segments\\& and CP tail groups\\
$\varepsilon_0$ & dielectric constant water\\
$\kappa$ & inverse Debye screening length \\
$\lambda = \frac{1}{2 \pi \sigma l_B}$ & Gouy-Chapman length \\
$\mu $ & Chemical potential CPs\\& in the presence of vRNA molecules\\
$\mu_{CP} $ & Chemical potential CPs\\& in the absence of vRNA molecules\\
$\xi_g$ & Correlation length or blob size of a\\& vRNA molecule in poor solvent\\
$\xi =l_B/b$ & Manning parameter, fraction \\ & of RNA charges not compensated \\ & by condensed counterions \\
$\Pi_{\psi} $ & Two dimensional surface pressure \\
$\rho_3$ & vRNA segment density \\
$\rho_2$ & number density of CPs \\ & on globule surface\\
$\sigma$ & surface charge density, \\ & in units of elementary charge $e$ \\
$\phi_{\mathrm{CP}}$ & solution concentration of\\& capsid proteins (CPs)\\
$\bar{\phi}_{\mathrm{CP}}$ & dimensionless solution concentration of\\& capsid proteins (CPs)\\
$\phi_{f}$ & solution concentration of\\& unbound capsid proteins (CPs)\\
$\bar{\phi_{f}}$ & dimensionless solution concentration of\\& unbound capsid proteins (CPs)\\
$\phi_{\mathrm{RNA}}$ & solution concentration of\\& vRNA molecules\\
$\bar{\phi}_{\mathrm{RNA}}$ & dimensionless solution concentration of\\& vRNA molecules\\
$\phi^{\ast}$ & critical protein concentration \\ & for virion assembly\\
$\psi$ & relative angle between the normals \\& of adjacent capsid proteins\\
$\psi_c$ & relative angle between the normals \\& of adjacent capsid proteins of the virion\\
$\Omega(R)$ & volume of a sphere of radius R\\

\hhline{|b:=:b:=:b|}
\end{tabular}
\end{table}


\end{document}